\documentclass[doublecol]{epl2} 
\usepackage[english]{babel}
\usepackage{amsmath,
amssymb,amsfonts,latexsym}
\usepackage{graphicx}
\usepackage{color}
\usepackage{subfigure}
\usepackage{afterpage}
\usepackage{booktabs}
\usepackage{blkarray}
\usepackage{cite}
\usepackage{ulem}
\usepackage{bm}
\usepackage{braket}
\usepackage{nicefrac}
\usepackage{etoolbox}
\usepackage{mathrsfs} 
\usepackage[protrusion=true,expansion=true]{microtype}
\usepackage{amsmath}
\usepackage[usenames,dvipsnames,table]{xcolor}
\usepackage[colorlinks,
citecolor=blue,linkcolor=red,urlcolor=blue]{hyperref}
\usepackage{amsfonts,amsmath,amssymb}
\usepackage{soul} 
\usepackage{enumitem} 
\usepackage{blindtext} 
\definecolor{Blue}{rgb}{0.00, 0.00, 1.00}
\definecolor{Red}{rgb}{1.00, 0.00, 0.00}
\definecolor{labelkey}{cmyk}{.1,.7,0.5,0}

\newcommand{\qqq}{\end{document}}

\include{new_commands}

\newcommand{\be}{\begin{equation}}
\newcommand{\ee}{\end{equation}}
\newcommand{\bea}{\begin{eqnarray}}
\newcommand{\eea}{\end{eqnarray}}

\title{Optimal mean first-passage time of a run-and-tumble particle in a class of one-dimensional confining potentials}
\shorttitle{Optimal MFPT of an RTP in a $1d$ confining potential}

\author{Mathis Gu\'eneau\inst{1} \and Satya N. Majumdar\inst{2} \and Gr\'egory Schehr \inst{1}}
\institute{\inst{1} Sorbonne Universit\'e, Laboratoire de Physique Th\'eorique et Hautes Energies, CNRS UMR 7589, 4 Place Jussieu, 75252 Paris Cedex 05, France\\
\inst{2} LPTMS, CNRS, Univ.  Paris-Sud,  Universit\'e Paris-Saclay,  91405 Orsay,  France
}
\shortauthor{}

\date{today}
\abstract{We consider a run-and-tumble particle (RTP) in one dimension, subjected to a telegraphic noise with a constant rate $\gamma$, and in the presence of an external confining potential $V(x) = \alpha |x|^p$ with $p \geq 1$. We compute the mean first-passage time (MFPT) at the origin $\tau_\gamma(x_0)$ for an RTP starting at $x_0$. We obtain a closed form expression for $\tau_\gamma(x_0)$ for all $p \geq 1$, which becomes fully explicit in the case $p=1$, $p=2$ and in the limit $p \to \infty$. For generic $p>1$ we find that there exists an optimal rate $\gamma_{\rm opt}$ that minimizes the MFPT and we characterize in detail its dependence on $x_0$. We find that $\gamma_{\rm opt} \propto 1/x_0$ as $x_0 \to 0$, while $\gamma_{\rm opt}$ converges to a nontrivial constant as $x_0 \to \infty$. In contrast, for $p=1$, there is no finite optimum and $\gamma_{\rm opt} \to \infty$ in this case. These analytical results are confirmed by our numerical simulations.     
}

\begin{document}

\maketitle

{\bf Introduction.} First-passage time (FPT) properties constitute a classical and notoriously difficult problem in the study of stochastic processes \cite{redner, BLMCV2011,Bray_review,Metzler_book}. During the last decades they have found myriad of applications ranging from chemical reaction kinetics \cite{Berg,Lindenberg2019}, intracellular transport \cite{BN2013} and slow dynamics in complex and disordered systems \cite{Bray_review,Satya_persistence} all the way to animals searching for foods \cite{BX2009,Viswanathan}, or more generally random search strategies \cite{BLMCV2011,Oshanin2007}. When the stochastic dynamics under study takes place in a confined geometry, such as in a bounded domain or in the presence of an external potential, the main features of the full distribution of the FPT are well captured by its average value, namely the mean first-passage time (MFPT) -- see e.g. \cite{Condamin2005}.

\begin{figure}[t]
    \centering
    \includegraphics[width=0.7\linewidth]{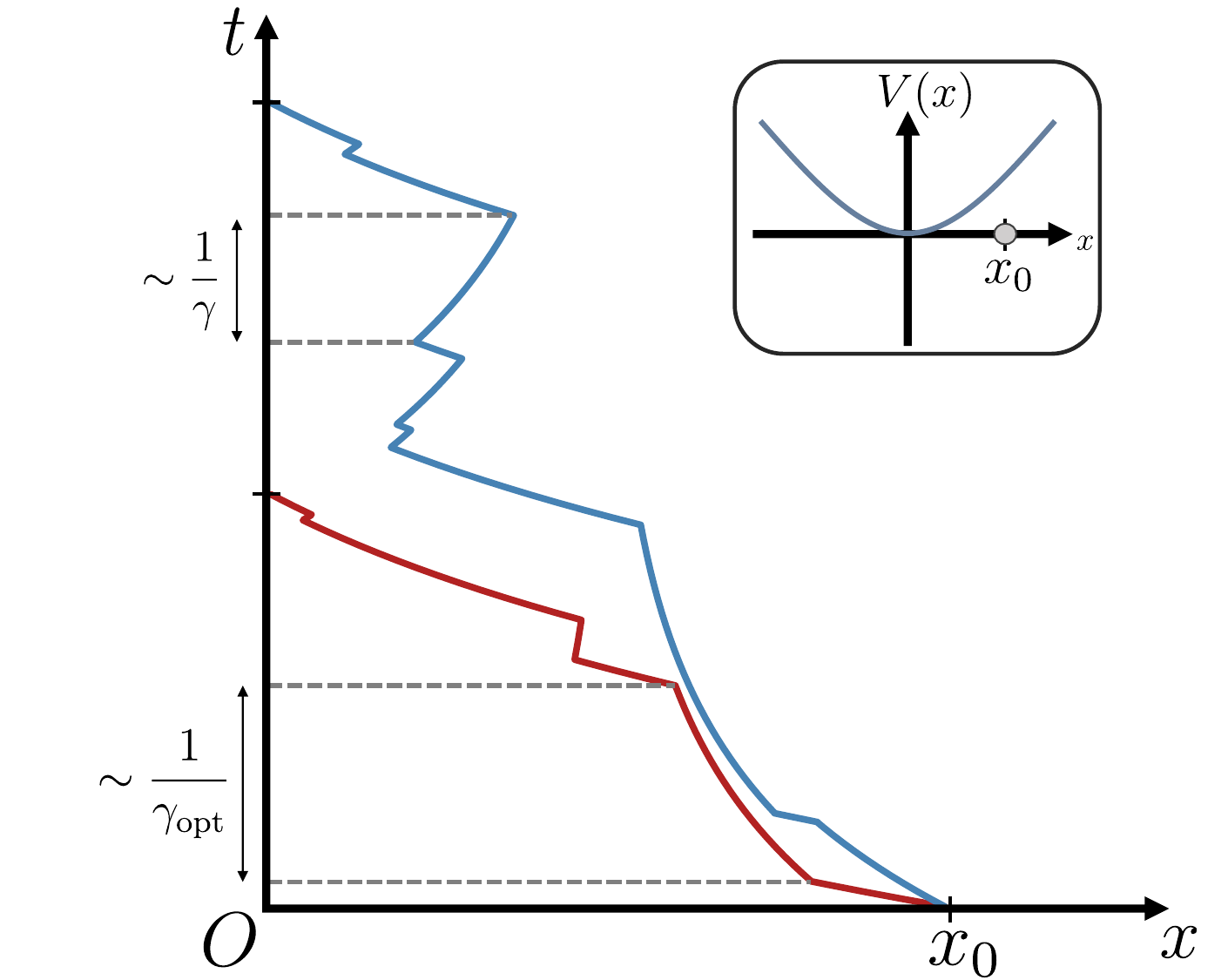}
  \caption{Schematic trajectories of an RTP in a confining potential $V(x)=\alpha |x|^p$ with $p>1$, for two different values of the tumbling rate $\gamma$. The typical time between two consecutive tumblings scales as $1/\gamma$. The trajectory stops when the particle hits the origin for the first time, starting at $x_0$. Here $\gamma_{\rm opt}$ denotes the optimal tumbling rate that minimizes the MFPT to the origin.}
  \label{figure_referee} 
\end{figure}

While the MFPT has been widely studied for passive particles, such as a Brownian particle in an external potential, much less is known for active particles. At variance with the passive particles, active particles are able to  consume energy from their environment, leading to self-propelled motion~\cite{Marchetti2013,Bechinger2016,Ramaswamy2017,Fodor2018}. In the absence of any confining potential, the first-passage properties of a single run-and-tumble active particle in one dimension has been computed exactly with many interesting results \cite{Masoliver1986,Weiss2002,Angelani2014,Angel2015,Malakar2018,PLD2019,Mori2020,Angel2023,EGK2023}. However, 
a central topic of current active matter research dwells on the effect of confinement, either in a finite system or in the presence of a confining trap (for example optical traps \cite{Dauchot,Takatori}). It is well known that, with increasing activity, the position distribution of an active particle undergoes a cross-over from a bell-shaped Gaussian distribution concentrated around the trap center to a structure with peaks near the edge of the trap signalling accumulation of the particles there \cite{Marchetti2013,Bechinger2016,Ramaswamy2017,Fodor2018,Dauchot,Takatori,Basu2019,RTPSS,Sevilla,Kardar}. In contrast, the first-passage properties of active particles in a confined geometry are much less explored. A quantity of prime interest is the MFPT, which has been studied for active particles in a finite domain, and in the absence of any potential, in dimensions up to $d=3$~\cite{TVB12,Angelani2014,Angel2015,RBV16,Malakar2018}. However, most experiments are performed in the presence of an optical trap, both harmonic and non-harmonic \cite{Dauchot,Takatori}. In Ref. \cite{RTPSS} the late time decay of the full first-passage probability of an RTP in a harmonic potential in $d=1$ was studied, though the MFPT is still unknown even in this case. The main goal of this paper is to derive exact analytical results for the MFPT of an RTP in $d=1$ in the presence of a variety of confining potentials of the form $V(x) = \alpha \, |x|^p$ with $p\geq 1$. The case $p=2$ corresponds to the harmonic potential. On the other hand, the limit $p \to \infty$ corresponds to an RTP in a finite domain $|x|<1$ with reflecting boundary conditions at $x=1$. Therefore our result makes a significant generalization of the existing results for the MFPT in a finite domain. 

One of our principal results here is that, for $p>1$,  there exists an optimal tumbling rate $\gamma$ that minimizes the MFPT. This has important implication for finding optimal strategies for active particles to navigate in a noisy environment \cite{Sano2017,Peleg2018,Lowen2019,Mori2023}. In fact, this question of optimizing the MFPT with respect to some parameters of the underlying dynamics is of paramount importance in many search strategies involving Brownian motion, L\'evy flights \cite{Stanley2011,LTBV2020}, stochastic resetting \cite{EM2011,KMSS2014,PR2017,Mercado, Zhang,EMS2020}, etc. For active particles, the optimization of MFPT has been studied mostly for systems in a finite domain~\cite{TVB12,RBV16,EM2018,BZBTV18,TGMS22,Angel2023, EGK2023,Paoluzzi,Ghosh,Debnath,Caprini}, but very little is known about such optimization in the presence of an external confining potential. Our exact results, in particular the existence of an optimal tumbling rate, valid for a wide class of one-dimensional potentials, will serve as important benchmarks for future studies of the MFPT in more complex environments.

A very popular model of active particles, inspired by the motion of {\it E. Coli} bacteria \cite{Berg_book}, is the so-called run-and-tumble particle (RTP) \cite{Weiss2002,ML2017} -- also called persistent random walk in the maths literature \cite{Kac1974,Ors1990} -- where the dynamics of the particle is subjected to a telegraphic (``active'') noise. In one-dimension, which we focus on here, the RTP evolves via the Langevin dynamics (see Fig. \ref{figure_referee})   
\begin{equation}
    \frac{dx(t)}{dt}=f(x) + v_0\, \sigma(t)\, ,
\label{langeRTP}
\end{equation}
where $f(x) = -V'(x)$ denotes the force acting on the RTP (and $V(x)$ the corresponding potential) and $v_0>0$ is a constant velocity. Here, $\sigma(t)$ is a telegraphic noise that takes values $\sigma(t) = \pm 1$ and alternates from one state to another with a constant rate $\gamma$. Hence, $\sigma(t)$ has zero mean, while its auto-correlation is given by $\langle \sigma(t) \sigma(t') \rangle = e^{-2 \gamma |t-t'|}$. This active noise is thus non-Markovian since it has a finite correlation/persistence time~$\gamma^{-1}$. The passive (or thermal) regime is recovered in the limit $\gamma \to \infty$, $v_0 \to \infty$ keeping $v_0^2/(2 \gamma) = D$ fixed, where the telegraphic noise $v_0 \sigma(t)$ converges to a Gaussian white noise with an associated diffusion coefficient $D$~(see e.g. \cite{Kac1974}).


A particularly interesting case corresponds to the class of confining potentials of the form $V(x) = \alpha\, |x|^p$, with $p \geq 1$. For this model, the position distribution $p_s(x)$ in the steady state is well known~\cite{RTPSS,Sevilla}. 
Indeed, $p_s(x)$ has actually a {\it finite} support $[-x_e,x_e]$ with two edges at $\pm x_e = \pm (v_0/(\alpha\,p) )^{1/(p-1)}$ which are the two fixed points of the dynamics (\ref{langeRTP}), i.e., $f(\pm x_e) = \mp v_0$. 
In addition, close to the edges at $\pm x_e$, the stationary distribution behaves as $p_s(x) \propto |x \mp x_e|^\phi$ with a nontrivial exponent $\phi$ given by
\bea \label{def_phi}
\phi = \frac{1}{p(p-1)} \left( \frac{v_0}{p}\right)^{\frac{2-p}{p-1}} \frac{\gamma}{\alpha^{1/(p-1)}} - 1 \;.
\eea 
Thus, for fixed $\alpha, v_0$ and $p$ there exists a critical value 
\bea \label{alpha_c}
\gamma_c = \gamma_c(p) = p(p-1)\alpha^{1/(p-1)}\left( \frac{p}{v_0}\right)^{\frac{2-p}{p-1}} \;,
\eea
such that $\phi < 0$  for $\gamma < \gamma_c$ while $\phi > 0$  for $\gamma > \gamma_c$~\footnote{In Ref. \cite{RTPSS}, the transition was described in the $(\alpha,p)$ plane, instead of $(\gamma,p)$, but it is of course totally equivalent.}. This indicates a shape transition for the distribution $p_s(x)$, from a bell shape with $\phi>0$ (the passive-like phase) to 
a U-shape with $\phi<0$ (the active-like phase where the particles accumulate at the edges)~\cite{RTPSS}. For $0<p<1$, the stationary distribution is trivially a delta-function centered at the origin and we will not discuss this case henceforth.

Given the solvable structure for the position distribution in this class of potentials, it is natural to ask whether the first-passage time distribution, or at least the MFPT, can be computed exactly for all $p\geq 1$. In Ref. \cite{RTPSS}, the Laplace transform of the first-passage distribution was computed only for the harmonic case $p=2$. However, even in this special case, extracting the MFPT from the Laplace transform is highly non-trivial and no explicit result is known.

In this Letter, we obtain analytical expressions for $\tau^{\pm}_\gamma(x_0)$, denoting the MFPT to the origin starting from the position $x_0 \geq 0$ in an initial state $\sigma(t=0) = \pm 1$, for any potential $V(x) = \alpha |x|^p$ with $p \geq 1$ and $\alpha>0$. For simplicity, we assume here that $\sigma(t=0) = \pm 1$ with equal probability $1/2$ such that 
the ``average'' MFPT is given by $\tau_\gamma(x_0) = (\tau_\gamma^{+}(x_0) + \tau_\gamma^-(x_0))/2$, which is given in Eq.~(\ref{phase3sol}) {for $p>1$. This expression in Eq.~(\ref{phase3sol}) becomes fully explicit in the case of a harmonic potential (i.e., $p=2$) -- see Eqs. (\ref{scaling_form})-(\ref{HarmonicscalingF}) -- as well as in the limiting case $p\to \infty$ -- see Eq. (\ref{tau_pinf}). Based on this exact formula~(\ref{phase3sol}), together with numerical simulations, we show that there indeed exists a finite optimal tumbling rate $\gamma_{\rm opt}$ that minimizes MFPT for all $p>1$, and we characterize its dependence on $x_0$ (see Fig. \ref{beta_optritic_figure} for $p=2$). Interestingly, we also find that the MFPT displays rather different behaviors for $p<2$ (left panel of Fig. \ref{MFPT_p_results}) and $p \geq 2$ (right panel of Fig. \ref{MFPT_p_results}). Indeed, while $\tau_\gamma(x_0)$ diverges as $\gamma \to 0$ for all $p>1$, it also diverges, as $\gamma \to \infty$, for $p\geq 2$ but remains finite for $1<p<2$. Below, we provide a physical explanation for these analytical results and numerical observations. We further show that only $\tau^+_\gamma(x_0)$ exhibits a minimum as a function of $\gamma$, while $\tau_\gamma^-(x_0)$ is a monotonically increasing function of $\gamma$.
Finally, we study the case $p=1$, i.e., $V(x) = \alpha |x|$, with $\alpha >0$, for which the MFPT can also be computed exactly. In this case, we show that there is no finite optimal rate $\gamma_{\rm opt}$ that minimizes $\tau_\gamma(x_0)$. Besides, we show that $\tau_\gamma(x_0)$ exhibits very different functional form for $\alpha < v_0$~(\ref{taulinearphase2a}) and $\alpha > v_0$~(\ref{linearalphalessv0}) -- see also Fig. \ref{MFPTLinearsimu1}.  
Establishing analytically the existence of an optimal tumbling rate $\gamma_{\rm opt}$ for all $p>1$ is one of the central results of this paper.



{\bf General framework}. We consider an RTP on the positive half-line, whose position is denoted by $x(t) \in [0,+\infty[$. We denote by $x(t=0)=x_0\geq0$ its initial position and it then evolves according to Eq. \eqref{langeRTP}. To compute the MFPT $\tau_\gamma^{\pm}(x_0)$, it is useful to introduce 
\begin{figure}[t]
    \centering
    \includegraphics[width=1\linewidth]{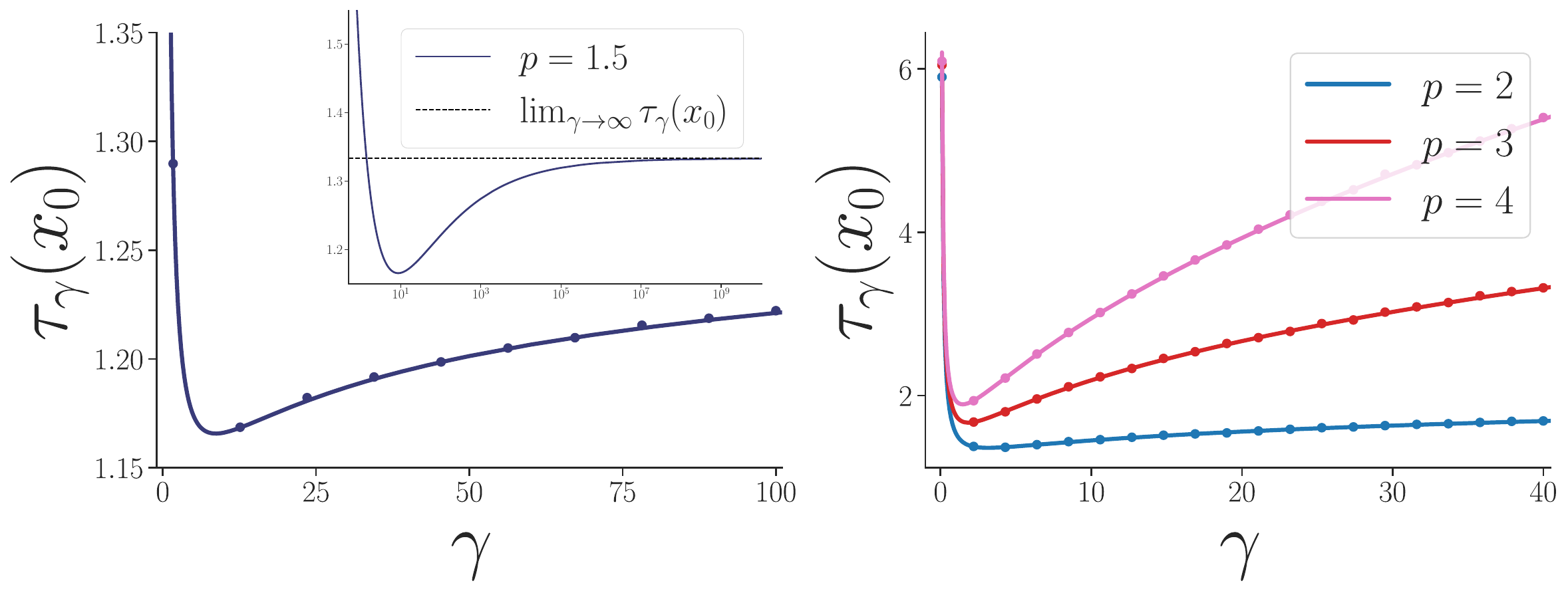}
  \caption{Plot of the MFPT $\tau_\gamma(x_0)$ as a function of the tumbling rate $\gamma$ for different values of $p$ (with $1<p<2$ on the left panel and $p\geq 2$ on the right panel). In both panels, the solid lines correspond to a numerical evaluation of our exact formula in Eq.~(\ref{phase3sol}) while the dots are the results of numerical simulations of (\ref{langeRTP}), showing an excellent agreement. In all these cases, there exists an optimal rate $\gamma_{\rm opt}$ that minimizes $\tau_\gamma(x_0)$. Here we used  $v_0 = 1$, $x_0=1$ and $\alpha=1$.}
  \label{MFPT_p_results} 
\end{figure}
the survival probabilities $Q^{\pm}(x_0,t)$, i.e., the probabilities that the position of the particle remains positive up to time $t$, starting from $x_0$ in the initial state $\sigma(0) = \pm 1$. By analysing the dependence of $Q^{\pm}(x_0,t)$ on the initial position $x_0$, one can show that they 
obey the following backward Fokker-Planck equations 
\cite{RTPSS,RTPPSG} 
\begin{eqnarray}
        \partial_t Q^+ = \left[f(x_0) + v_0\right]\partial_{x_0} Q^+ -\gamma\, Q^+ + \gamma\, Q^-\, , \label{survivalequations1}\\
        \partial_t Q^- = \left[f(x_0) - v_0\right]\partial_{x_0} Q^- -\gamma\, Q^- + \gamma\, Q^+\, .\label{survivalequations2}
\end{eqnarray}
In particular, the ``average'' survival probability of the RTP is $Q(x_0,t)=(Q^+(x_0,t)+Q^-(x_0,t))/2$. From the coupled equations~(\ref{survivalequations1}) and~(\ref{survivalequations2}), it is straightforward to write the corresponding differential equations 
for the mean first passage times $\tau_\gamma^{\pm}(x_0)$. Indeed, we first simply notice that the distribution of the first-passage time $F(x_0,t)$ is given by $F(x_0,t)=-\partial_t Q(x_0,t)$ \cite{redner,Bray_review}, implying $\tau^{\pm}_\gamma(x_0)  = -\int_0^{\infty}dt\, t\, \partial_t Q(t,x_0)$. Therefore, by taking one time derivative of Eqs.~(\ref{survivalequations1}) and~(\ref{survivalequations2}), multiplying them by $t$ and integrate over $t$ (using $Q^{\pm}(x_0>0,t=0) = 1$), we obtain~\cite{RTPPSG} 
%
\begin{eqnarray}
    && \left[f(x_0)+ v_0\right] \partial_{x_0}\tau^+_\gamma - \gamma \tau^+_\gamma + \gamma \tau^-_\gamma= -1 \, ,\label{coupledtaupm1}\\
&&\left[f(x_0) - v_0\right] \partial_{x_0}\tau^-_\gamma + \gamma \tau^+_\gamma -\gamma \tau^-_\gamma  = -1\, ,
\label{coupledtaupm2}
\end{eqnarray}
which are valid for $x_0 >0$. From Eqs.~(\ref{coupledtaupm1}) and (\ref{coupledtaupm2}), it is possible to obtain a second order ordinary differential equation for $\tau_\gamma= \left(\tau^+_\gamma(x_0) + \tau^-_\gamma(x_0)\right)/2$, which reads~\cite{SM} 
\begin{eqnarray}
    &&\left[v_0^2 -f(x_0)^2\right] \partial^2_{x_0} \tau_\gamma+ 2f(x_0)\left[\gamma-f'(x_0) \right]\partial_{x_0} \tau_\gamma\nonumber\\
    &&= f'(x_0)- 2\gamma\, .
\label{ODE2ndTau}
\end{eqnarray}
Once $\tau_\gamma(x_0)$ is known,  $\tau^\pm_\gamma(x_0)$ can then be obtained from~\cite{SM} 
\begin{equation}
\tau^-_\gamma(x_0) = \frac{f(x_0)}{2\gamma\, v_0}- \frac{1}{2\gamma v_0}\left[v_0^2-f(x_0)^2\right]\partial_{x_0} \tau_\gamma+ \tau_\gamma\, ,\label{ODEtauminus}
\end{equation}
\begin{equation}
\tau^+_\gamma(x_0) = - \frac{f(x_0)}{2\gamma\, v_0}+  \frac{1}{2\gamma v_0}\left[v_0^2-f(x_0)^2\right]\partial_{x_0} \tau_\gamma+ \tau_\gamma\, .\label{ODEtauplus}
\end{equation}
To specify a unique solution of (\ref{ODE2ndTau}), we need to fix two integration constants. {We will treat the case $p=1$ separately at the end of the letter, and concentrate for now on $p>1$.} We first notice that for $f(x) = -\alpha p |x|^{p-1}$,} the MFPT of an RTP that starts its motion at the origin with $\sigma(0)=-1$ is zero since the particle crosses the origin immediately (indeed the force acting on it $f(0)-v_0 = -v_0 < 0$). Thus $\tau^-_\gamma(x_0=0) = 0$, while $\tau^+_\gamma(x_0=0)$ remains unspecified and in general is nonzero. 
%
\begin{figure}[t]
    \centering
    \includegraphics[width=0.795\linewidth]{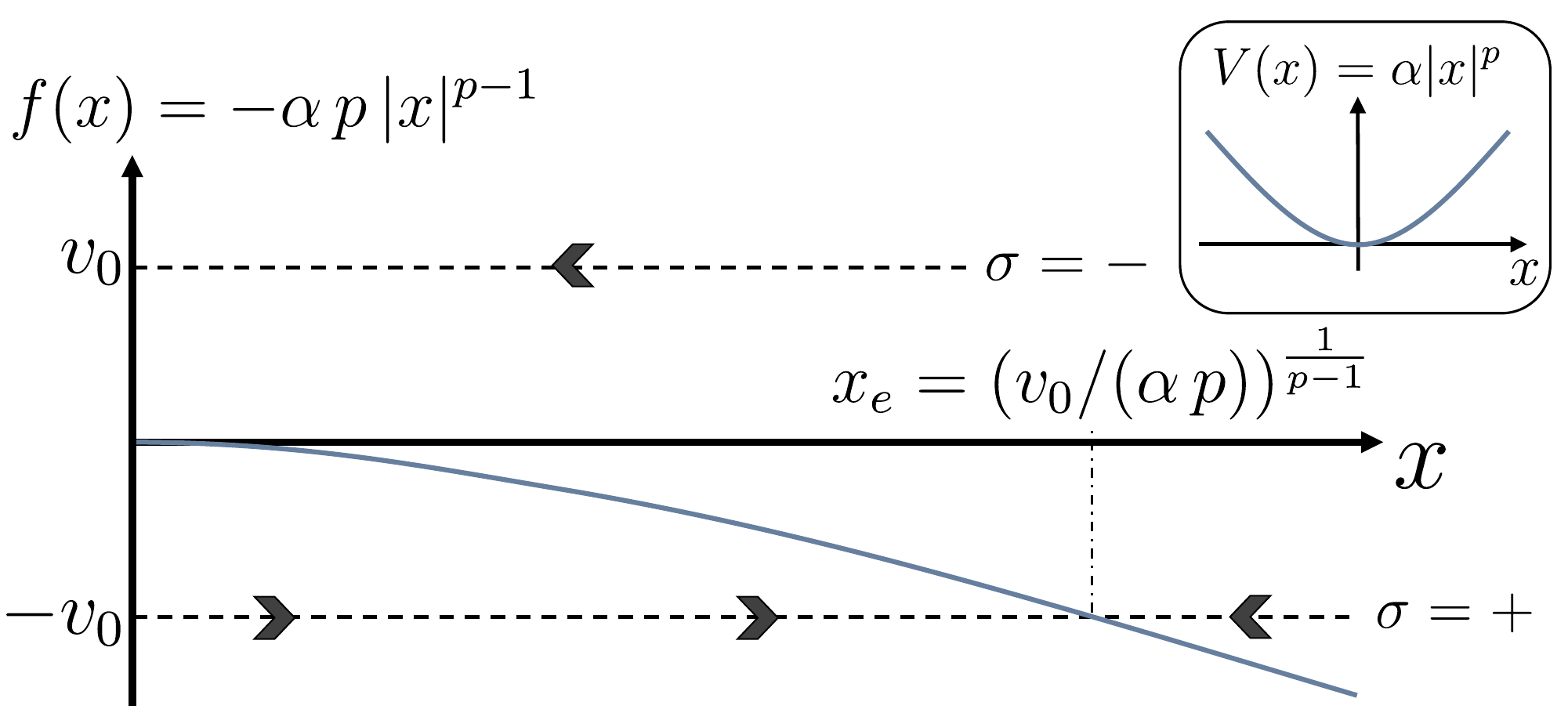}
  \caption{Schematic motion of an RTP inside a confining potential $V(x)=\alpha |x|^p$, and the associated force is $f(x)=-\alpha\, p\, x^{p-1}$, when $p>1$. The force has a stable negative fixed point $x_e= (v_0/(\alpha\, p))^{\frac{1}{p-1}}$ such that $f(x_e)=-v_0$. Note that when $1<p\leq2$, the concavity of the force is not the same as the one represented in the figure. The arrows on the dotted lines show the direction of the velocity of the RTP in state $\sigma=\pm$. If the arrow is directed to the right (left), the velocity is positive (negative) in this region.}
  \label{MFPTconfiningforceFig} 
\end{figure}

To obtain a second boundary condition, let us analyse the case where the RTP starts from the stable fixed point at $x_0=x_e$, with $f(x_e)=-v_0$ in the initial state '$+$'. In the time interval $[0, dt]$, the position of the particle evolves according to
\begin{equation} \label{rule_dt}
x(dt) = \begin{cases}
x_e&\text{, w. proba. } 1 - \gamma\, dt \text{, } \sigma(dt) = +1 \\
x_e - 2 v_0 \,dt&\text{, w. proba. } \gamma\, dt \text{, } \sigma(dt) = -1
\end{cases}\, ,
\end{equation}
where the first line corresponds to the case where $\sigma$ remains unchanged (i.e., positive) while the second line applies if $\sigma$ has flipped from '+' to '-'. One can then write a backward equation for $Q^+(x_0,t+dt)$ by decomposing the time interval into $[0,dt]$ and $[dt,t+dt]$. Using the evolution during the time interval $[0,dt]$ in (\ref{rule_dt}), one clearly has
\begin{equation} \label{BFP_Q+}
Q^+(x_e,t+dt) = Q^+(x_e,t) (1-\gamma dt) + Q^-(x_e - 2 v_0 dt,t) \gamma dt \;.
\end{equation}
Expanding for small $dt$ one finds 
\begin{eqnarray}
    \partial_t Q^+(x_e,t) =  -\gamma\, Q^+(x_e,t) + \gamma\, Q^-(x_e,t)\, .
\label{BFP_xe}
\end{eqnarray}
Therefore, to ensure the compatibility between Eq. (\ref{survivalequations1}) and Eq. (\ref{BFP_xe}), we need to impose the second condition
\begin{eqnarray} \label{cond2_1}
  \lim_{x_0 \to x_e}  \left[\left(f(x_0)+v_0\right)\partial_{x_0}Q^+(x_0,t)\right]=0\, \;.
\end{eqnarray}
As before, taking one time derivative of (\ref{cond2_1}), multiplying by $t$ and integrating it over $t$, one obtains  
\begin{eqnarray} \label{cond2}
     \lim_{x_0 \to x_e} \left[\left(f(x_0)+v_0\right)\partial_{x_0}\tau^+_\gamma(x_0)\right]=0\, .
\end{eqnarray}
By using Eqs. \eqref{ODEtauminus} and \eqref{ODEtauplus}, it is easy to see that the same condition (\ref{cond2}) applies also to $\tau_\gamma(x_0)$ \cite{SM}. Therefore, for a force $f(x) = - \alpha \, p\, x^{p-1}$ with one unique stable negative fixed point $f(x_e)=-v_0$, we conclude that $\tau_\gamma(x_0)$ is the unique solution of 
the differential equation (\ref{ODE2ndTau}) satisfying two following conditions
\begin{equation}
    \tau^-_\gamma(x_0=0) = 0 \, ,
    \label{condition1}
\end{equation}
\begin{equation}
   \lim_{x_0 \to x_e} \left[\left(f(x_0)+v_0\right)\partial_{x_0}\tau_\gamma(x_0)\right]=0\, ,
   \label{condition2}
\end{equation}
where we recall that $\tau_\gamma^-(x_0)$ is given by Eq. (\ref{ODEtauminus}). Under these conditions, Eq. (\ref{ODE2ndTau}) can be solved explicitly \cite{SM}
 \begin{equation}
    \begin{split}
     &\tau_\gamma(x_0) =   \frac{1}{2\gamma}+\int_{0}^{x_e} \frac{dy}{v_0-f(y)} \, \frac{1}{H(y)} + \\
     &\int_0^{x_0}dz\, \frac{1}{v_0^2-f(z)^2} \int_{x_e}^{z} dy \left(f'(y)- 2\gamma \right) \frac{H(z)}{H(y)}\, ,
    \end{split}
    \label{phase3sol}
\end{equation} 
where $H(x) = \text{exp}\left[\int_0^{x}du\, \frac{-2\gamma f(u)}{v_0^2-f(u)^2}\right]$. The argument in the exponential can be interpreted as an effective potential, since in the passive limit $\gamma \to \infty$, $v_0 \to \infty$, keeping $v_0^2/(2 \gamma) = D$ fixed, $H(x) \to \exp{[(V(x)-V(0))/D]}$~\cite{RTPPSG}. In fact, by inserting this limiting form in Eq. (\ref{phase3sol}), one finds that in the passive limit only the term in the second line of Eq. (\ref{phase3sol}) survives, leading to 
\begin{eqnarray} \label{passive_tau}
\tau_\gamma(x_0) \sim \frac{1}{D} \int_0^{x_0} dz \, \int_z^{\infty} dy \exp{\left(\frac{V(z)-V(y)}{D} \right)} \;,
\end{eqnarray}
which coincides with the known formula for passive systems, see e.g. \cite{Blatter98}.


This formula (\ref{phase3sol}) holds actually for a wide class of force field $f(x)$ such that the equation $f(x) = -v_0$ has a single root $x_e$ with $f'(x_e) <0$ (and $f(x) < v_0$ \cite{us_tbp}). In the specific case $f(x) = - \alpha p x^{p-1}$, it turns out that $\tau_\gamma(x_0)$ can be written under the scaling form~
\begin{eqnarray}\label{scaling_form}
\tau_\gamma(x_0) = \frac{1}{\gamma_c} {\cal F}_p \left(\beta = \frac{\gamma}{\gamma_c}, u = \frac{x_0}{x_e} \right) \;,   
\end{eqnarray} 
where we recall that $x_e = (v_0/(\alpha\,p) )^{1/(p-1)}$ while $\gamma_c$ is given in Eq. (\ref{alpha_c}). For general $p>1$, the MFPT $\tau_\gamma(x_0)$ (or equivalently ${\cal F}_p(\beta, u)$) is given in terms of a double integral, which can however be studied in detail (see below). However for $p=2$, on which we first focus, $\tau_\gamma(x_0)$ can be explicitly computed.


{\bf The case $p=2$ (harmonic potential)}. In this case where $V(x) = \mu x^2/2$ (i.e., we use $\alpha = \mu/2$), 
the double integral in Eq. (\ref{phase3sol}) can be computed explicitly and $\tau_\gamma(x_0)$ takes the scaling form as in Eq. (\ref{scaling_form}) with $\gamma_c = \mu$, $x_e = v_0/\mu$ and 
%
\begin{eqnarray}
&       \mathcal{F}_2\left(\beta,u\right) = \frac{\sqrt{\pi}}{2\beta} \frac{\Gamma\left(1+\beta\right)}{\Gamma\left(\frac{1}{2}+\beta\right)}\left[1 + 2\beta\, u \, {}_2 F_1\left(\frac{1}{2},1 +\beta,\frac{3}{2};u^2\right)\right] \nonumber\\
&-  (2\beta + 1) \, \frac{u^2}{2} \, {}_3 F_2\left(\{1,1,\frac{3}{2}+\beta\},\{\frac{3}{2},2\};u^2\right)\,,
\label{HarmonicscalingF}
\end{eqnarray}
where ${}_2 F_1(\cdot;z)$ and ${}_3 F_2(\cdot;z)$ are hypergeometric functions~\cite{Grad}. For some special values of $\beta$, it takes a simpler form. For instance ${\cal F}_2(1,u) = 2-1/(1+u)+\log(1+u)$. Note that although these two hypergeometric series have a branch cut along $[1,+\infty)$, the combination of these two that enters the expression of ${\cal F}_2(\beta,u)$ in Eq. (\ref{HarmonicscalingF}) is, for any $\beta > 0$, a perfectly smooth function of $u$ on the whole real line, and in particular at $u=1$ (i.e., for $x$ close to the edge $x_e$). We recall that one can straightforwardly deduce $\tau^\pm_\gamma$ from (\ref{scaling_form}) and (\ref{HarmonicscalingF}) using (\ref{ODEtauminus}) and (\ref{ODEtauplus}). 
In what follows we focus on the optimal value $\gamma_{\rm opt}$ associated to $\tau_\gamma(x_0)$ (but a similar analysis can be carried out for $\gamma^+_{\rm opt}$ associated to $\tau^+_\gamma(x_0)$).  

From the explicit expression (\ref{HarmonicscalingF}) one can straightforwardly obtain the behaviors of $\tau_\gamma(x_0)$ in the two opposite limits $\gamma \to 0$ and $\gamma \to \infty$. They read (returning back to dimensionful variables)
\begin{eqnarray}
  \tau_{\gamma}(x_0) \sim 
\begin{cases}
\dfrac{1}{2\gamma} \quad {\rm ,} \quad \gamma \to 0\\
\\
\frac{1}{2 \gamma_c}\, \log (\gamma/\gamma_c) \quad {\rm ,} \quad \gamma \to +\infty
\end{cases}
\, .
\label{var_asymp_harmonic_gamma}
\end{eqnarray}
{Interestingly, the behavior of $\tau_\gamma(x_0)$ in the two limits $\gamma \to 0$ and $\gamma \to \infty$ are independent of $x_0$.}
The divergence in the two limits are easy to understand physically. When $\gamma \to 0$, the RTP is highly persistent and if it starts with a positive velocity at $x_0 > 0$ it will converge to the fixed point at $+x_e$ (see Fig. \ref{MFPTconfiningforceFig}) and it will get stuck there for a very long time before the velocity reverses direction (since $\gamma \to 0$). This makes the MFPT divergent as $\gamma \to 0$. In the opposite limit $\gamma \to \infty$, the telegraphic noise behaves as white noise with an effective diffusion constant $D=v_0^2/(2 \gamma)$. Thus the diffusion constant of the effective Ornstein-Uhlenbeck (OU) particle also vanishes in this limit for fixed $v_0$. Consequently the MFPT for the effective OU process also diverges as $\gamma \to \infty$, which can be checked from Eq. (\ref{passive_tau}) when $D\to 0$. Since $\tau_\gamma(x_0)$ diverges in these two limits, this shows that there exists an optimal value $\gamma_{\rm opt}$, which we can reasonably assume to be unique. It is however difficult to compute $\gamma_{\rm opt}$ for any finite $x_0$. To 
make progress we analyse the two limits $x_0 \to 0$ and $x_0 \to \infty$. For small $x_0$, the MFPT $\tau_\gamma(x_0)$ behaves as 
\begin{eqnarray} \label{small_x0}
\tau_\gamma(x_0) \approx \frac{1}{\gamma_c}\left(A_\gamma + B_\gamma \frac{x_0}{x_e} \right) \;,
\end{eqnarray}
where $A_\gamma = \sqrt{\pi} \, \Gamma(\frac{\gamma}{\gamma_c})/\Gamma(\frac{1}{2} + \frac{\gamma}{\gamma_c})$ and $B_\gamma = 2 (\gamma/\gamma_c) A_\gamma$. By analysing the $\gamma$-dependence of the coefficient $A_\gamma$ and $B_\gamma$, one finds that there indeed exists an optimal value $\gamma_{\rm opt}$ which is diverging as $x_0 \to 0$. Expanding $A_\gamma$ and $B_\gamma$ for large $\gamma$ one finds that $\gamma_{\rm opt}$ behaves, when $x_0 \to 0$, as
\begin{eqnarray} \label{opt_small_p2}
\gamma_{\rm opt} \approx \frac{\gamma_c}{2} \frac{x_e}{x_0} \;.
\end{eqnarray}

For large $x_0$, it is also straightforward to obtain the behavior of $\tau_\gamma(x_0)$ from Eqs. (\ref{scaling_form}) and (\ref{HarmonicscalingF}). It reads 
\begin{eqnarray}
\tau_\gamma(x_0) = \frac{1}{\gamma_c} \left(\log (x_0/x_e) + c_\gamma + o(1) \right) \;,
\end{eqnarray}
where $c_\gamma$ is a constant (i.e., independent of $x_0$) that can be computed explicitly. Here we see that only the subleading term, namely the constant $c_\gamma$ depends on $\gamma$. We find that $c_\gamma$ indeed admits a minimum for $\gamma_{\rm opt} = \beta^* \, \gamma_c $ where $\beta^* = 1.38657 \ldots$ is the solution of a transcendental equation~\cite{SM}.   
To summarize, the optimal dimensionless rate $\beta_{\rm opt}=\gamma_{\rm opt}/\gamma_c$ , as a function of the scaled initial distance $u=x_0/x_e$, behaves asymptotically as
\begin{eqnarray}
\beta_{\rm opt} = \frac{\gamma_{\rm opt}}{\gamma_c} \sim 
\begin{cases}
\dfrac{x_e}{2 x_0} \quad {\rm ,} \quad \hspace*{2cm} u \to 0 \;,\\
\\
\beta^* =1.38657\dots\, \quad {\rm ,} \quad u \to +\infty
\end{cases}
\, .
\label{beta_opt_harmonic_asympt}
\end{eqnarray}
In Fig. \ref{beta_optritic_figure} we show a plot of $\beta_{\rm opt}$ as a function of $u=x/x_e$ which has been obtained from the numerical minimization with respect to $\beta$ (for different values of $u=x/x_e$) of the exact expression in Eq. (\ref{HarmonicscalingF}). We also compare this curve with the exact asymptotic behaviors in Eq. (\ref{beta_opt_harmonic_asympt}), showing a very good agreement in both limits. 


\begin{figure}[t]
    \centering
    \includegraphics[width=0.7\linewidth]{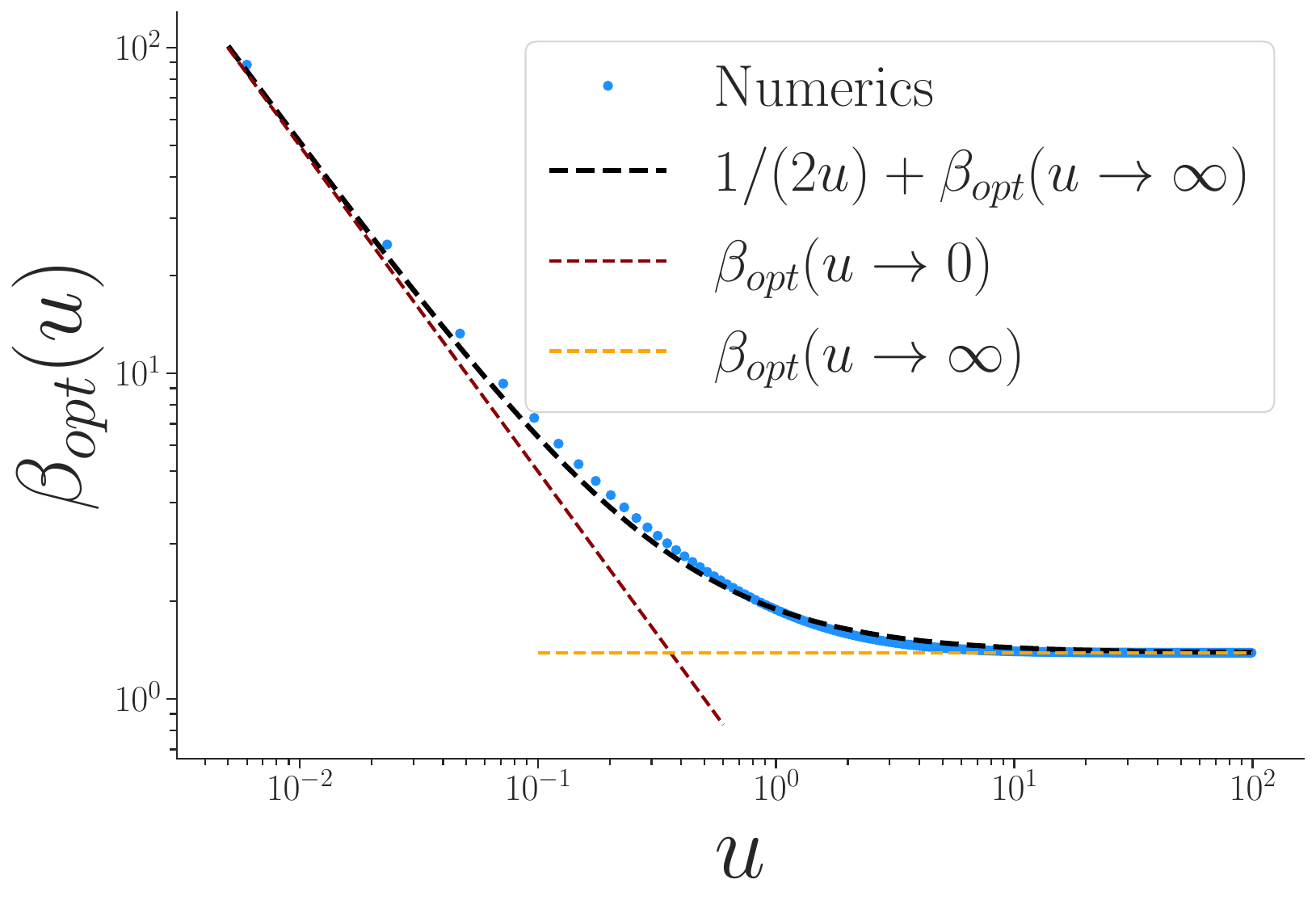}
  \caption{Plot of $\beta_{\rm opt} = \gamma_{\rm opt}/\gamma_c$ for the harmonic potential $V(x)=\mu\, x^2/2$ as a function of $u = x_0/x_e = \mu x_0/v_0$ on a log-log scale. The blue data points represent numerical computations of the minimum using Eq.~(\ref{HarmonicscalingF}) while the red and yellow dotted lines correspond to the asymptotic behaviors of $\beta_{\rm opt}$ for small and large $u$ as given in Eq.~(\ref{beta_opt_harmonic_asympt}). The black dotted line is the sum of the two asymptotic behaviors, which describes quite accurately the whole curve.}
  \label{beta_optritic_figure} 
\end{figure}

{\bf The generic case $p>1$}. Let us now turn to the generic case. From the formula (\ref{phase3sol}), one can extract the asymptotic behaviors of $\tau_\gamma(x_0)$ in the two limits $\gamma \to 0$ and $\gamma \to \infty$. In the limit $\gamma \to 0$, one finds \cite{SM} 
\begin{eqnarray} \label{small_gamma}
     \tau_\gamma(x_0) \approx \frac{1}{2\gamma}  \;,
\end{eqnarray}
independently of $p>1$. The mechanism responsible for this divergence as $\gamma \to 0$ is again due to the long persistence time $1/\gamma$ of the RTP, as discussed in the $p=2$ case below Eq. (\ref{var_asymp_harmonic_gamma}).
The large $\gamma$ behavior of $\tau_\gamma(x_0)$ is a bit more subtle. It can again be obtained from Eq. (\ref{phase3sol}) and we find that, depending on $p$ it exhibits different behaviors, namely \cite{SM}
\begin{eqnarray} \label{asympt_large_gamma}
\tau_\gamma(x_0) \underset{\gamma \to \infty}{\sim}\begin{cases}
\frac{x_0^{2-p}}{\alpha p(2-p)} \quad {\rm ,} \quad 1<p<2 \\
\\
A_p\, \gamma^{1 -\frac{2}{p}} \quad  {\rm ,} \quad p>2
\end{cases}
\, ,
\label{var_asymp_letter}
\end{eqnarray}
where $A_p$ is a computable constant \cite{SM}. We recall that exactly at $p=2$, we found a logarithmic behavior [see Eq. (\ref{var_asymp_harmonic_gamma})]. These behaviors in (\ref{small_gamma}) and (\ref{var_asymp_letter}) are in agreement with the numerical results shown in Fig. \ref{MFPT_p_results}. The limit $\gamma \to \infty$ corresponds to passive motion with a diffusion coefficient $D = v_0^2/(2\gamma) \to 0$, since $v_0$ is finite here. In this limit, Eq. (\ref{langeRTP}) thus reduces to the standard Langevin equation but in the absence of noise, i.e., $dx/dt = f(x)$, since $D \to 0$. Hence the MFPT is just the time to reach the origin, starting from $x_0$, which reads in this limit $\tau_\gamma(x_0) = \lim_{\epsilon\to 0} - \int_\epsilon^{x_0} dx/f(x)$. For $p<2$, this limit is well defined and, for $f(x) = - \alpha p x^{p-1}$ reproduces exactly the first line of Eq. (\ref{asympt_large_gamma}). On the other hand, for $p>2$, the integrand $1/f(x)$ has a nonintegrable singularity at the origin and therefore $\tau_{\gamma = 0}(x_0)$ diverges, which is consistent with the second line of (\ref{asympt_large_gamma}). Note that these behaviors (\ref{asympt_large_gamma}) can also be recovered by analysing the passive formula (\ref{passive_tau}) in the limit $D \to 0$ limit. 

Therefore, for $p>2$, the MFPT $\tau_\gamma(x_0)$ diverges in both limits $\gamma \to 0$ and $\gamma \to \infty$ (see the right panel of Fig. \ref{MFPT_p_results}), which shows that there is an optimal rate $\gamma_{\rm opt}$. Nevertheless, this analysis is not conclusive for $p<2$ since $\tau_\gamma(x_0)$ remains finite for $\gamma \to \infty$ (see the left panel of Fig. \ref{MFPT_p_results}). To make further progress, we considered the limit $x_0 \to 0$ where we could show that there is an optimal $\gamma_{\rm opt}$ for all $p>1$ given by the compact formula (see~\cite{SM})
\begin{eqnarray}\label{gamma_opt_smallx0}
\gamma_{\rm opt} \approx \frac{\gamma_c}{2(p-1)^2} \frac{x_e}{x_0} \;, \;\; {\rm as} \;\; x_0 \to 0 \;,
\end{eqnarray}
which matches with the result in Eq.~(\ref{opt_small_p2}) for $p=2$. Similarly, one can also analyse the limit $x_0 \to \infty$ and show that there is an optimal tumbling rate $\gamma_{\rm opt}$ which converges to a $p$-dependent constant in that limit~\cite{SM}.

{\bf The case $p \to \infty$.} In this limiting case, one can show that the function $H(x)$ in Eq. (\ref{phase3sol}) reduces simply to 
$H(x) = 1$ for all $x \geq 0$. Hence the integrals in Eq. (\ref{phase3sol}) can be performed explicitly. Besides $x_e \to 1$ in this limit and, interestingly, one finds that $\tau_\gamma(x_0)$ has a well defined limit as $p\to \infty$ which reads
\begin{eqnarray} \label{tau_pinf}
\tau_\gamma(x_0) =
\begin{cases}
&\frac{1}{2\gamma} + \frac{x_0+1}{v_0}+ \frac{\gamma x_0(2-x_0)}{v_0^2} \;, \;x_0\leq 1 \;, \\
&\frac{1}{2\gamma} + \frac{2}{v_0}+ \frac{\gamma}{v_0^2} \;, \quad \quad\quad\;\;\, \;\; x_0 > 1 \;.
\end{cases}
\end{eqnarray}
It is easy to see from Eq. (\ref{tau_pinf}) that there exists an optimal rate $\gamma_{\rm opt}$ that minimizes $\tau_\gamma(x_0)$. This optimal rate reads
\begin{eqnarray} \label{opt_pinf}
\gamma_{\rm opt} =
\begin{cases}
&\frac{v_0}{\sqrt{2x_0(2-x_0)}} \quad, \quad x_0 \leq 1 \;, \\
&\frac{v_0}{\sqrt{2}} \quad, \quad x_0 > 1 \;.
\end{cases}
\end{eqnarray}
Hence in this case, at variance with the case $p=2$ (see Fig.~\ref{beta_optritic_figure}), $\gamma_{\rm opt}$ exhibits a transition at $x_0 = 1$ where its second derivative with respect to $x_0$ is discontinuous, while $\gamma_{\rm opt}$ itself and its first derivative are continuous. We note that the result for $x_0<1$ in (\ref{opt_pinf}) can also be obtained by noticing that in the limit $p \to \infty$ the potential $V(x) = \alpha |x|^p$ is equivalent (for particle coming from $x_0<1$) to a reflecting hard wall at $x=1$. It does indeed exhibit an optimal $\gamma_{\rm opt}$ and this result in $d=1$ is thus the counterpart of the ones found for $d=2,3$ in Refs.~\cite{TVB12,RBV16}.

{\bf The special case $p=1$}. It turns out that for a linear potential $V(x) = \alpha\, |x|$ where $\alpha >0$, the stationary state is different from the case $p>1$ \cite{RTPSS}. In this case, indeed, there is a nontrivial stationary state $p_{s}(x)$ only for $\alpha < v_0$, while $p_{s}(x) = \delta(x)$ for $\alpha >v_0$. In addition, for $\alpha < v_0$, at variance with the case $p>1$, the support of $p_s(x)$ is the whole real line, consistent with the fact that $x_e \to +\infty$ as $p \to 1$. It is thus natural to ask if there is a signature of this critical value $\alpha_c=v_0$ in the MFPT and if there still exists an optimal rate $\gamma_{\rm opt}$ in this case.

By analysing Eq. (\ref{ODE2ndTau}), it is easy to see that the two aforementioned cases $\alpha < v_0$ and $\alpha > v_0$ need to be treated separately. 
For $0<\alpha <v_0$, we find that $\tau_\gamma(x_0)$ is given by Eq.~(\ref{phase3sol}) with the limit $x_e \to \infty$  
\begin{figure}[t]
    \centering
    \includegraphics[width=1\linewidth]{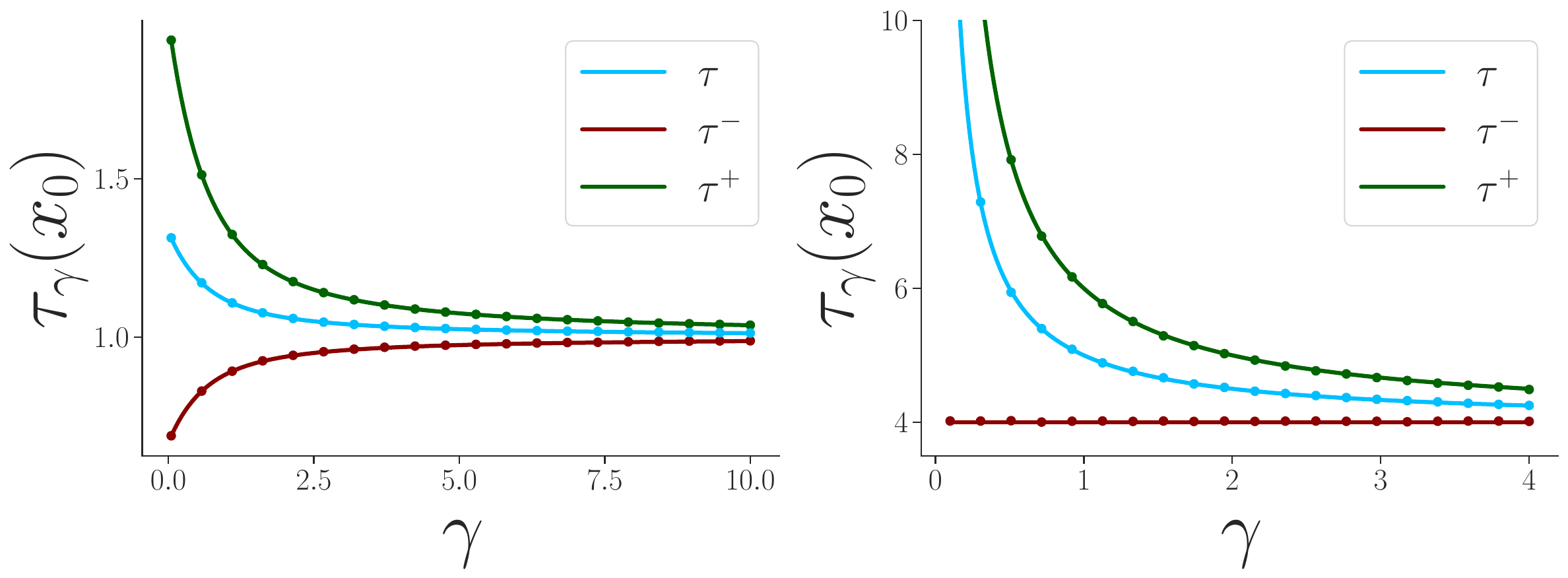}
  \caption{{\bf Left panel:} Plot of the MPFT's for $V(x) = \alpha|x|$ ($p=1$) for $\alpha < v_0$ as a function of $\gamma$. Here $v_0=0.5$, $\alpha = 1$ and $x_0 = 1$. {\bf Right panel:} Same plot but for $\alpha > v_0$ -- here $v_0=1$, $\alpha = 2$ and $x_0=1$. In both panels, the solid lines are the exact analytical results in Eqs. (\ref{taulinearphase2a}) and (\ref{linearalphalessv0}) while the points are the results of numerical simulations. At variance with the case $p>1$ in Fig. \ref{MFPT_p_results}, there is no finite $\gamma_{\rm opt}$ for $p=1$.}
  \label{MFPTLinearsimu1} 
\end{figure}
(in fact, it is equivalent to the situation with a hard wall at position $L$ when $L \to \infty$ as discussed above \cite{SM}). We obtain
\begin{eqnarray}
  \tau_\gamma(x_0) =  \frac{x_0}{\alpha} + \frac{v_0}{2 \alpha \gamma} \;, \; \alpha < v_0 \;.
  \label{taulinearphase2a}
\end{eqnarray}
From the knowledge of $\tau_\gamma(x_0)$, one can also compute $\tau^\pm_\gamma(x_0)$ using Eqs.~(\ref{ODEtauminus}) and (\ref{ODEtauplus}).
In the left panel of Fig. \ref{MFPTLinearsimu1}, we show results of simulations and compare them to the analytical results, showing a very good agreement. {Interestingly, when $\alpha < v_0$ the stationary distribution $p_s(x)$ is a double-exponential \cite{RTPSS} akin to a free RTP under resetting~\cite{EM2018}. By matching the rate parameters of the distributions, one could compare the MFPT for the two models to see if one outperforms the other, as in the diffusive case \cite{diff_vs_reset} (see~\cite{us_tbp}). }

In the opposite case $\alpha>v_0$, since $f(0)=-\alpha < -v_0$, the initial velocity in the two states of the RTP $\dot{x}(0)=-\alpha \pm v_0$ are both negative. This implies 
\begin{eqnarray}
\tau_\gamma^+(x_0=0) = 0\; , \; \tau^-_\gamma(x_0=0) = 0 \;. \label{conditiondtau2bLinear}
\end{eqnarray}
Because of different boundary conditions, we find that the expression of $\tau_\gamma(x_0)$ is not given by the general expression in (\ref{phase3sol}) which holds for $p>1$. Instead, by solving explicitly  Eq.~(\ref{ODE2ndTau}) with (\ref{conditiondtau2bLinear}) one finds 
\begin{equation}
\tau_\gamma(x_0) =  \frac{x_0}{\alpha}+ \frac{v_0^2}{2 \alpha^2 \gamma}\left(1- e^{-\frac{2\alpha \gamma x_0}{\alpha^2-v_0^2}}\right)\, \;, \quad \alpha > v_0 \;.
\label{linearalphalessv0}
\end{equation}
Again, using Eq.~(\ref{ODEtauminus}) and Eq.~(\ref{ODEtauplus}), we can compute the MFPT's $\tau_\gamma^{\pm}(x_0)$. 
For a comparison with simulations, see the right panel of Fig. \ref{MFPTLinearsimu1}.


The first observation is that for $p=1$, there is no finite $\gamma_{\rm opt}$ that minimizes the MFPT (see Fig. \ref{MFPTLinearsimu1}): this is in sharp contrast with the case $p>1$ studied above. In addition, by considering $\tau_\gamma(x_0)$ in Eqs. (\ref{taulinearphase2a}) and (\ref{linearalphalessv0}) as a function of $\alpha$, we see that, although it is continuous as $\alpha$ crosses $v_0$, the first derivative of $\tau_\gamma(x_0)$ with respect to $\alpha$ is discontinuous at $\alpha = v_0$. 
It shows that when $p=1$ the MFPT carries a signature of the transition found in the stationary state.
This is in contrast to the case $p>1$ where the transition from a bell-shape to a U-shape does not show up so drastically in the MFPT, except in the limit $p \to \infty$.

{\bf Conclusion.} In summary, we have studied in detail the MFPT at the origin $\tau_\gamma(x_0)$ for an RTP starting from
$x_0 >0$ with a tumbling rate $\gamma$ and in the presence of an external confining potential $V(x) = \alpha |x|^p$ with $p \geq 1$. We have obtained a closed form expression for $\tau_\gamma(x_0)$, valid for all $p>1$. From this result we have shown that there exists an optimal tumbling rate $\gamma_{\rm opt}$ which minimizes the MFPT. We also characterized in detail the dependence of $\gamma_{\rm opt}$ as a function of $x_0$. For generic $p>1$ it diverges as $\gamma_{\rm opt}\propto 1/x_0$ for small $x_0$ and converges to a constant as $x_0 \to \infty$. However, this scenario does not hold for $p=1$, for which there is no finite optimal rate, i.e. $\gamma_{\rm opt} \to \infty$. 

Of course, it is natural to ask what happens to more general potentials, e.g., in the case $p<1$. Another interesting future direction would be to study the MFPT of an RTP in higher dimensions~\cite{SLMS2022,Smith2023}. For example, in $d=2$, it would be interesting to consider an anisotropic harmonic potential $V(x,y) = \mu_1 x^2/2 + \mu_2 y^2/2$. Here we have studied the limiting case $\mu_2 \to \infty$ where the problem becomes effectively one-dimensional. We expect our results to be valid (up to some renormalization of time scales) when $\mu_2$ is finite but $\mu_2 \gg \mu_1$, while the results for the isotropic case (when $\mu_2 = O(\mu_1)$) may differ significantly and it would be interesting to explore this, since this is also relevant for experiments in optical traps.

{\it Acknowledgments:} we thank P. Le Doussal and L. Touzo for useful discussions and ongoing collaborations on related subjects.

{}

\newpage 

\onecolumn

\begin{center}
{\large{\bf Supplementary material for {\it Optimal mean first-passage time of a run-and-tumble particle in a one-dimensional confining potential} by M. Gu\'eneau, S. N. Majumdar and G. Schehr}}
\end{center}

\vspace*{0.5cm}

{In this Supplementary material, we give the derivation of some results given in the main text.}

\vspace*{0.5cm}

\begin{center}
{\bf I. Derivation of the differential equation of the MFPT $\tau_\gamma$, and the expressions of $\tau_\gamma^\pm$}
\end{center}

In this Section, we derive the differential equation for $\tau_\gamma$ as well as the expression of $\tau_\gamma^\pm$ with respect to $\tau_\gamma$ and its first derivative, given respectively in Eqs. (8), (9) and (10) in the main text. As 
recalled in the text, and derived in \cite{RTPPSG}, $\tau_\gamma^{\pm}(x_0)$ obey the following coupled differential equations 
\begin{eqnarray}
    && \left[f(x_0)+ v_0\right] \partial_{x_0}\tau_\gamma^+(x_0) - \gamma \tau_\gamma^+(x_0) + \gamma \tau_\gamma^-(x_0)= -1 \, ,\label{coupledappendix2}\\
&&\left[f(x_0) - v_0\right] \partial_{x_0}\tau_\gamma^-(x_0) + \gamma \tau_\gamma^+(x_0)  -\gamma \tau_\gamma^-(x_0)  = -1\, .\label{coupledappendix1}
\end{eqnarray}
The first goal is to derive an ordinary differential equation (ODE) for the MFPT $\tau_\gamma(x_0)=1/2\left(\tau_\gamma^+(x_0) +\tau_\gamma^-(x_0) \right)$. 
Equations (\ref{coupledappendix2}) and (\ref{coupledappendix1}) can be also written in terms of operators $\mathcal{L}_\pm$ acting on $\tau_\gamma^\pm$ as
\begin{eqnarray}
    && \mathcal{L}_+\tau_\gamma^+(x_0)= \left[\left(f(x_0)+ v_0\right)\partial_{x_0} - \gamma\right] \tau_\gamma^+(x_0) =-1- \gamma \tau_\gamma^-(x_0) \, ,\label{coupledappendix2bis}\\
&&\mathcal{L}_-\tau_\gamma^-(x_0)=\left[\left(f(x_0)- v_0\right)\partial_{x_0} - \gamma\right] \tau_\gamma^-(x_0) =-1 - \gamma \tau_\gamma^-(x_0) \, .\label{coupledappendix1bis}
\end{eqnarray}
Applying $\mathcal{L}_-$ on the left hand side (LHS) of Eq.~(\ref{coupledappendix2bis}) and using Eq.~(\ref{coupledappendix1bis}) gives a second order differential equation on $\tau_\gamma^+$. A similar procedure where one applies $\mathcal{L}_+$ on the LHS of Eq.~(\ref{coupledappendix1bis}) instead gives the equation for $\tau_\gamma^-$. It gives
\begin{equation}
\left[v_0^2 -f(x_0)^2\right] \partial^2_{x_0} \tau_\pm(x_0) + \left[2\gamma f(x_0) - f(x_0)f'(x_0) \pm v_0f'(x_0)\right]\partial_{x_0} \tau_\pm(x_0) = -2\gamma\, .
\end{equation}
Next, by summing up these two equations for $\tau_\gamma^+$ and $\tau_\gamma^-$, one finds
\begin{equation} \label{steptau1}
2\left[v_0^2 -f(x_0)^2\right] \partial^2_{x_0} \tau_\gamma(x_0) + 2\left[2\gamma f(x_0) - f(x_0)f'(x_0)\right]\partial_{x_0} \tau_\gamma(x_0) + v_0f'(x_0)\partial_{x_0} \left(\tau_\gamma^+(x_0) -\tau_\gamma^-(x_0)\right) = -4\gamma\, .
\end{equation}
We now want to express the difference $ \left(\tau_\gamma^+(x_0) -\tau_\gamma^-(x_0)\right)$ in terms of the MFPT $\tau_\gamma(x_0)$. To this end, one can sum the two equations (\ref{coupledappendix2}) and (\ref{coupledappendix1}). We obtain
\begin{equation}\label{differencetaus}
\partial_{x_0} \left(\tau_\gamma^+(x_0) -\tau_\gamma^-(x_0)\right) = -\frac{2}{v_0}\left(1+f(x_0)\partial_{x_0}\tau_\gamma(x_0)\right)\, .
\end{equation} 
One can then insert Eq.~(\ref{differencetaus}) in Eq. (\ref{steptau1}) and simplify to obtain the ODE on $\tau_\gamma(x_0)$
\begin{equation}\label{generalequationontau}
\left[f(x_0)^2- v_0^2 \right] \partial^2_{x_0} \tau_\gamma(x_0) + 2f(x_0)\left[f'(x_0) -\gamma\right]\partial_{x_0} \tau_\gamma(x_0) = 2\gamma -f'(x_0)\, ,
\end{equation}
which is Eq. (8) given in text. 

In addition, from the knowledge of $\tau_\gamma(x_0)$, it is possible to obtain the full expressions of $\tau_\gamma^-(x_0)$ and $\tau_\gamma^+(x_0)$. By definition we have
\begin{equation}
\tau_\gamma(x_0) = \frac{1}{2}\left(\tau_\gamma^+(x_0) + \tau_\gamma^-(x_0)\right) \, ,
\label{deftauappendix}
\end{equation}
so that
\begin{equation}
\tau_\gamma^+(x_0) = 2\tau_\gamma(x_0) - \tau_\gamma^-(x_0)\, .
\label{tau_+wrttautauminus}
\end{equation}
Substituing this relation in Eq.~(\ref{differencetaus}) gives
\begin{equation}
\partial_{x_0} \left(2\tau_\gamma(x_0) -2\tau_\gamma^-(x_0)\right) = -\frac{2}{v_0}\left(1+f(x_0)\partial_{x_0}\tau_\gamma(x_0)\right)\, ,
\end{equation} 
which can also be written as
\begin{equation}
\partial_{x_0} \tau_\gamma^-(x_0) = \left(1+\frac{f(x_0)}{v_0}\right)\partial_{x_0}\tau_\gamma(x_0)  + \frac{1}{v_0}\, .
\label{toolcoupledtaupm}
\end{equation} 
Substituting Eqs. (\ref{tau_+wrttautauminus}) and (\ref{toolcoupledtaupm}) in Eq. (\ref{coupledappendix1}) yields an expression for $\tau_\gamma^-(x_0)$ in terms of $\tau_\gamma(x_0)$ and $\partial_{x_0} \tau_\gamma(x_0)$. And using again  Eq. (\ref{tau_+wrttautauminus}) one finally obtains 
\begin{eqnarray}
    &\tau_\gamma^-(x_0) = \frac{1}{2\gamma}\frac{f(x_0)}{v_0} - \frac{v_0}{2\gamma}\left(1-\frac{f(x_0)^2}{v_0^2}\right)\partial_{x_0} \tau_\gamma(x_0) + \tau_\gamma(x_0) \label{tauminuswrttau}\\
&\tau_\gamma^+(x_0) = - \frac{1}{2\gamma}\frac{f(x_0)}{v_0} +  \frac{v_0}{2\gamma}\left(1-\frac{f(x_0)^2}{v_0^2}\right)\partial_{x_0} \tau_\gamma(x_0) + \tau_\gamma(x_0) \, ,\label{taupluswrttau}
\end{eqnarray}
as given in Eqs. (9) and (10) in the main text.

\begin{center}
{\bf II. Explicit solution of the ODE satisfied by $\tau_\gamma(x_0)$}
\end{center}

In this section, we will derive the mean first passage time (MFPT) $\tau_\gamma(x_0)$ for $x_0\in [0,+\infty[$ when the force has a unique stable negative turning point $x_e$ that satisfies $f(x_e)=-v_0$, and $f(x_0) <v_0$ (as considered in the main text).

Let us solve the differential equation for the MFPT $\tau_\gamma(x_0)$. We recall that the MFPT obeys
\begin{equation}
\left[v_0^2 -f(x_0)^2\right] \partial^2_{x_0} \tau_\gamma + 2f(x_0)\left[\gamma-f'(x_0) \right]\partial_{x_0} \tau_\gamma = f'(x_0)- 2\gamma\, .
\end{equation}
We introduce $w(x_0)=\partial_{x_0}\tau_\gamma$ such that the differential equation satisfied by $w(x_0)$ is of first order, namely
\begin{equation}
\left[v_0^2 -f(x_0)^2\right] \partial_{x_0}w(x_0) + 2f(x_0)\left[\gamma-f'(x_0) \right]w(x_0) = f'(x_0)- 2\gamma\, .
\label{eq1storder}
\end{equation}
One can first solve the homogeneous equation, whose solution is denoted by $w_H$, i.e.,  
\begin{equation}
\left[v_0^2 -f(x_0)^2\right] \partial_{x_0}w_H(x_0) + 2f(x_0)\left[\gamma-f'(x_0) \right] w_H(x_0) = 0\, .
\end{equation}
The solution is then
\begin{equation} \label{wH}
    w_H(x_0)= \frac{A}{v_0^2-f(x_0)^2}H(x_0)\quad , \quad {\rm where} \quad  H(x_0) = \text{exp}\left[\int_0^{x_0}dx \frac{-2\gamma f(x)}{v_0^2-f(x)^2}\right] \;,
\end{equation} 
and $A$ is an integration constant. 
The solution of the full equation (\ref{eq1storder}) can then be obtained from (\ref{wH}), e.g., by varying the constant, and one finds
%
\begin{equation}
    w(x_0) = \frac{H(x_0)}{v_0^2-f(x_0)^2}\left[\int_0^{x_0} dy \frac{f'(y)- 2\gamma}{H(y)} + A\right]\, .
\label{wsolwithA}
\end{equation}

In the main text we have shown that when the force has a {unique} stable negative fixed point $x_e$, and when $f(x)<v_0$, the boundary conditions are [see Eqs. (16) and (17) in the main text]
\begin{equation}
    \tau_\gamma^-(x_0=0) = 0 \, ,
    \label{condition1appendix}
\end{equation}
\begin{equation}
   \lim_{x_0 \to x_e} \left(f(x_0)+v_0\right)\partial_{x_0}\tau_\gamma^+(x_0)=0\, .
\label{conditionTauplus}
\end{equation}
First, let us show that condition~(\ref{conditionTauplus}) on $\tau_\gamma^+$ also applies to $\tau_\gamma$. Using the same procedure that was used to derive Eq.~(\ref{toolcoupledtaupm}), it is easy to show that we also have
\begin{eqnarray}
      \partial_{x_0}\tau_\gamma^+ = \left(1-\frac{f(x_0)}{v_0}\right)\partial_{x_0}\tau_\gamma - \frac{1}{v_0}\, .
\end{eqnarray}
As by definition $f(x_e)=-v_0$, the condition~(\ref{conditionTauplus}) also imposes
\begin{equation}
   \lim_{x_0 \to x_e} \left(f(x_0)+v_0\right)\partial_{x_0}\tau_\gamma(x_0)=0\, .
\label{conditionTauappendix}
\end{equation}

Now, we will use the condition~(\ref{conditionTauappendix}) to fix $A$, and we recall that $w(x_0)=\partial_{x_0}\tau_\gamma$. Hence (\ref{conditionTauappendix}) simply reads
\begin{equation}
   \lim_{x_0 \to x_e} \left(f(x_0)+v_0\right)w(x_0)=0\, .
   \label{conditiononw}
\end{equation}
To fix $A$, we need to analyse $w(x_0)$ in (\ref{wsolwithA}) close to the fixed point $x_e$. Let us write $f(x_e+\epsilon) = -v_0 + \epsilon f'(x_e)$. First we study the behavior of the function $H(x_0)$ in Eq. (\ref{wH}). For simplicity, we take the logarithm of $H$,
\begin{eqnarray}
    \ln \left(H(x_0)\right) =\int_0^{x_0}dx \frac{-2\gamma f(x)}{v_0^2-f(x)^2}\, .
\end{eqnarray}
Taking one derivative with respect to (w.r.t.) $x_0$ one finds
\begin{eqnarray}
\partial_{x_0} \left[ \ln \left(H(x_0)\right)\right] \Big \vert_{x_0 = x_e + \epsilon} = 
    \frac{-2\gamma f(x_e+\epsilon)}{v_0^2-f(x_e+\epsilon)^2} \sim \frac{\gamma}{\epsilon f'(x_e)}\, \;.
\end{eqnarray}
Integrating back w.r.t. to $x_0$ one finds
\begin{eqnarray}
    \ln \left(H(x_0=x_e +\epsilon)\right) \approx \frac{\gamma}{f'(x_e)}\ln|\epsilon| + \text{constant}\, ,
\end{eqnarray}
such that
\begin{eqnarray} \label{asympt_H}
    H(x_e+\epsilon) \sim |\epsilon|^{\frac{\gamma}{f'(x_e)}} = |\epsilon|^{-\frac{\gamma}{|f'(x_e)|}}\, ,
\end{eqnarray}
where we have used the fact that $x_e$ is here a {\it stable} fixed point, i.e., $f'(x_e) < 0$. 
It is now possible to analyse the integral inside the bracket in Eq.~(\ref{wsolwithA}) close to $x_e$. We have
\begin{eqnarray} \label{exp_int}
    \int_0^{x_0} dy \frac{f'(y)- 2\gamma}{H(y)} =  \int_0^{x_e} dy \frac{f'(y)- 2\gamma}{H(y)} + \int_{x_e}^{x_0} dy \frac{f'(y)- 2\gamma}{H(y)}\, ,
\end{eqnarray}
where the first integral on the right hand side (RHS) is perfectly well defined, thanks to the behavior in (\ref{asympt_H}). 
Therefore the expansion of this integral (\ref{exp_int}) for $x_0$ close to $x_e$ reads, to leading order
\begin{eqnarray} \label{exp_int2}
    \int_0^{x_e + \epsilon} dy \frac{f'(y)- 2\gamma}{H(y)} \approx  \int_0^{x_e} dy \frac{f'(y)- 2\gamma}{H(y)} + \text{sign}(\epsilon)\, |\epsilon|^{1+\frac{\gamma}{|f'(x_e)|}}\left(f'(x_e)-2\gamma\right)\, .
\end{eqnarray}
By inserting this behavior in the expression for $w(x_0)$ in Eq. (\ref{wsolwithA}) one thus finds
%
\begin{equation} \label{exp_int3}
  \left(f(x_e+ \epsilon)+v_0\right)w(x_e+ \epsilon) \propto |\epsilon|^{\frac{-\gamma}{|f'(x_e|)}}\left[\int_0^{x_e} dy \frac{f'(y)- 2\gamma}{H(y)} +A+ \text{sign}(\epsilon)\, |\epsilon|^{1+\frac{\gamma}{|f'(x_e)|}}\left(f'(x_e)-2\gamma\right)\right]\, .
\end{equation}
Therefore, to fulfill the condition (\ref{conditiononw}), we need to impose
\begin{eqnarray}
    A = -\int_0^{x_e} dy \frac{f'(y)- 2\gamma}{H(y)}\, .
\end{eqnarray}
Note that the higher order terms in $\epsilon$ are all analytic in (\ref{exp_int3}), which indicates that $\tau_\gamma(x_0)$ is a perfectly smooth function close to $x_e$. Therefore $w(x_0)$ is given by
\begin{equation}
    w(x_0) = \frac{H(x_0)}{v_0^2-f(x_0)^2} \int_{x_e}^{x_0} dy \frac{f'(y)- 2\gamma}{H(y)}\, .
\label{wexpressionfinal}
\end{equation}
The MFPT is then given by integrating Eq.~(\ref{wexpressionfinal}) over $x_0$, i.e., 
\begin{equation}
    \tau_\gamma(x_0) = B +
    \int_0^{x_0}dz\, \frac{H(z)}{v_0^2-f(z)^2} \int_{x_e}^{z} dy \frac{f'(y)- 2\gamma}{H(y)}\, .
\label{}
\end{equation}
To fix $B$, we use condition~(\ref{condition1appendix}), i.e., $\tau_\gamma^-(0) = 0$. Using Eq.~(\ref{tauminuswrttau}) we have
\begin{equation}
    \tau_\gamma^-(x_0) = B+ \frac{1}{2\gamma}\frac{f(x_0)}{v_0} + \frac{H(x_0) }{2\gamma v_0}\int_{x_0}^{x_e} dy \frac{f'(y)- 2\gamma}{H(y)} -\int_0^{x_0}dz\, \frac{H(z)}{v_0^2-f(z)^2} \int_{z}^{x_e} dy \frac{f'(y)- 2\gamma}{H(y)}\, .
\end{equation} Hence,\begin{equation}
    B = - \frac{1}{2\gamma}\frac{f(0)}{v_0} - \frac{1 }{2\gamma v_0}\int_{0}^{x_e} dy \frac{f'(y)- 2\gamma}{H(y)}\, .
\end{equation}
Finally, we have  
\begin{equation}
    \tau_\gamma(x_0) =  - \frac{1}{2\gamma v_0}\left[f(0) + \int_{0}^{x_e} dy \frac{f'(y)- 2\gamma}{H(y)}\right]  +\int_0^{x_0}dz\, \frac{H(z)}{v_0^2-f(z)^2} \int_{x_e}^{z} dy \frac{f'(y)- 2\gamma}{H(y)}\, \;.
\label{tau_FP_1}
\end{equation}
Using some further manipulations explained below, it is possible to rewrite the first term, corresponding actually to $\tau_\gamma(0)$, in a more compact way, given in  Eq.~(\ref{T_01}). This finally leads to 
the result given in Eq. (18) in the text.  

\begin{center}
{\bf III. Asymptotic behaviors of the MFPT} 
\end{center}

The starting point of our analysis is Eq. (18) of the main text, namely 
\begin{equation}
    \begin{split}
     \tau_\gamma(x_0) &=   \frac{1}{2\gamma}+\int_{0}^{x_e} \frac{dy}{v_0-f(y)} \, \text{exp}\left[\int_0^{y}du\, \frac{2\gamma f(u)}{v_0^2-f(u)^2}\right]  \\
     &+\int_0^{x_0}dz\, \frac{1}{v_0^2-f(z)^2} \int_{x_e}^{z} dy \left(f'(y)- 2\gamma \right) \text{exp}\left[\int_y^{z}du\, \frac{-2\gamma f(u)}{v_0^2-f(u)^2}\right]\, .
    \end{split}
    \label{phase3solappendix}
    \end{equation}
In this section, we study the asymptotic behaviors of Eq.~(\ref{phase3solappendix}) as a function of $\gamma$ and $x_0$ when the force derives from a potential $V(x) = \alpha|x|^p$, with $p>1$. We thus have $f(x)=-\alpha\, p\, x^{p-1}$ and  $x_e= (v_0/(\alpha\, p))^{\frac{1}{p-1}}$. It is convenient to introduce 
\begin{eqnarray} \label{I1}
    I_1(\gamma) = \int_{0}^{x_e} \frac{dy}{v_0-f(y)} \, \text{exp}\left[\int_0^{y}du\, \frac{2\gamma f(u)}{v_0^2-f(u)^2}\right]\, ,
\end{eqnarray}
as well as
\begin{eqnarray}
    I_2(x_0,\gamma) = \int_0^{x_0}dz\, \frac{1}{v_0^2-f(z)^2} \int_{x_e}^{z} dy \left(f'(y)- 2\gamma \right) \text{exp}\left[\int_y^{z}du\, \frac{-2\gamma f(u)}{v_0^2-f(u)^2}\right]\;, \label{I2} 
\end{eqnarray}
such that
\begin{eqnarray}
    \tau_\gamma(x_0) = \frac{1}{2\gamma} + I_1(\gamma) + I_2(x_0, \gamma)\, .
\end{eqnarray}
  
\vspace*{0.3cm}
 \begin{center} 
{\bf A.} {\it The limit $\gamma \to 0$}
\end{center}

The small $\gamma$ limit is quite straightforward. In fact, the arguments inside the exponentials go to $0$ and it is easy to show that we have
\begin{eqnarray}
    \tau_\gamma(x_0) \approx \frac{1}{2\gamma} + \int_0^{x_e}\, \frac{dy}{v_0-f(y)}  + \int_0^{x_0}\, \frac{dz}{v_0-f(z)} \quad \, , \quad \gamma \to 0\, .
\end{eqnarray}
In particular, when $p=2$, and $\alpha = \mu/2$ it gives
\begin{eqnarray}
     \tau_\gamma(x_0) \approx \frac{1}{2\gamma} + \frac{1}{\mu}\left[\ln 2 +\ln\left(1+\frac{\mu\, x_0}{v_0}\right)\right] \quad \, , \quad \gamma \to 0\, ,
\end{eqnarray}
which coincides with the direct expansion of the expression given in Eq. (19) in the main text in terms of hypergeometric functions. 
In any case,  the leading term for small $\gamma$ is simply
\begin{eqnarray}\label{smallgammabehaviorappendix}
     \tau_\gamma(x_0) \approx \frac{1}{2\gamma}  \quad \, , \quad \gamma \to 0\, ,
\end{eqnarray}
as given in the first line of Eq. (21) in the text. Hence the MFPT diverges when $\gamma \to 0$.

\vspace*{0.3cm}

\begin{center}
{\bf B.} {\it The limit $\gamma \to \infty$}
\end{center}

When $\gamma$ is large, the analysis is more subtle. Let us start by rewriting the function $H(x)$ in Eq. (\ref{wH}) explicitly (with $f(x) = - \alpha p x^{p-1}$)
\begin{eqnarray}
    H(x) = \text{exp}\left[-\int_0^{x}du\, \frac{2\gamma f(u)}{v_0^2-f(u)^2}\right] = \text{exp}\left[\int_0^{x}du\, \frac{2\gamma \, \alpha\, p\, u^{p-1}}{v_0^2-\alpha^2\, p^2\, u^{2p-2}}\right]\,
\end{eqnarray}
With the change of variable $\tilde u = \gamma^{\frac{1}{p}}\, u$, it reads
\begin{eqnarray}
    H(x) =  \text{exp}\left[\int_0^{x \, \gamma^{\frac{1}{p}}}d\tilde u\, \frac{2\, \alpha\, p\, \tilde u^{p-1}}{v_0^2-\frac{\alpha^2\, p^2\, \tilde u^{2p-2}}{\gamma^{\frac{1}{p}(2p-2)}}}\right]\, .
\end{eqnarray}

The first integral $I_1(\gamma)$ in (\ref{I1}) reads
\begin{eqnarray}
  I_1(\gamma) =   \int_{0}^{x_e} \frac{dy}{v_0-f(y)} \, \frac{1}{H(y)} =\int_{0}^{x_e} \frac{dy}{v_0+\alpha\, p\, y^{p-1}} \, \frac{1}{H(y)}\, .
\end{eqnarray}
Again, one can perform the change of variable $\tilde y = \gamma^{\frac{1}{p}}\, y$ to obtain
\begin{eqnarray}
  I_1(\gamma) = \frac{1}{\gamma^{\frac{1}{p}}} \int_{0}^{x_e\, \gamma^{\frac{1}{p}}} \frac{d\tilde y}{v_0+\frac{\alpha\, p\, \tilde y^{p-1}}{\gamma^{\frac{1}{p}(p-1)}}} \, \text{exp}\left[-\int_0^{\tilde y}d\tilde u\, \frac{2\, \alpha\, p\, \tilde u^{p-1}}{v_0^2-\frac{\alpha^2\, p^2\, \tilde u^{2p-2}}{\gamma^{\frac{1}{p}(2p-2)}}}\right]\, .
\end{eqnarray}
When $\gamma \to +\infty$, as $p>1$, we have
\begin{eqnarray}
    I_1(\gamma) \approx \frac{1}{\gamma^\frac{1}{p}}\, \int_0^{+\infty}\frac{d\tilde y}{v_0}\, e^{-\frac{2\alpha}{v_0^2}\tilde y^p}  \quad \, , \quad \gamma \to + \infty\, .
    \label{I1changedlimit}
\end{eqnarray}

Now for the second integral $I_2(x_0,\gamma)$ in (\ref{I2}) we apply three change of variables: $\tilde u = \gamma^{\frac{1}{p}}\, u$, $\tilde y = \gamma^{\frac{1}{p}}\, y$ and $\tilde z = \gamma^{\frac{1}{p}}\, z$. It gives
\begin{eqnarray}
    I_2(x_0,\gamma) = \gamma^{-\frac{2}{p}} \int_0^{x_0\, \gamma^{\frac{1}{p}}}d\tilde z\, \frac{1}{v_0^2-\frac{\alpha^2 p^2 \tilde z^{2p-2}}{\gamma^{\frac{1}{p}(2p-2)}}} \int_{\tilde z}^{x_e\, \gamma^{\frac{1}{p}}} d\tilde y \left(\frac{\alpha p (p-1)\tilde y^{p-2}}{\gamma^{\frac{1}{p}(p-2)}}+2\gamma \right) \text{exp}\left[\int_{\tilde y}^{\tilde z}d\tilde u\, \frac{2\alpha \, p \, \tilde u^{p-1}}{v_0^2-\frac{\alpha^2 p^2 \tilde u^{2p-2}}{\gamma^{\frac{1}{p}(2p-2)}}}\right]\, .
    \label{I2changed}
\end{eqnarray}
If $p>1$, when $\gamma \to \infty$ one finds simply
\begin{eqnarray}
    I_2(x_0,\gamma) \approx \frac{2\, \gamma^{1 -\frac{2}{p}}}{v_0^2} \int_0^{x_0\, \gamma^{\frac{1}{p}}}d\tilde z \int_{\tilde z}^{+\infty} d\tilde y \, e^{\frac{2\alpha}{v_0^2}(\tilde z^p - \tilde y^p)}  \quad \, , \quad \gamma \to + \infty\, .
\label{I2changedlimit}
\end{eqnarray}
Hence, at large $\gamma$, the MFTP reads 
\begin{eqnarray}
    \tau_\gamma(x_0) \approx \frac{1}{2\gamma} + \frac{1}{\gamma^\frac{1}{p}}\, \int_0^{+\infty}\frac{d\tilde y}{v_0}\, e^{-\frac{2\alpha\, p}{v_0^2}\tilde y^p} +\frac{2\, \gamma^{1 -\frac{2}{p}}}{v_0^2} \int_0^{x_0\, \gamma^{\frac{1}{p}}}d\tilde z \int_{\tilde z}^{+\infty} d\tilde y \, e^{\frac{2\alpha}{v_0^2}(\tilde z^p - \tilde y^p)}  \quad \, , \quad \gamma \to + \infty\, .
\end{eqnarray}
In particular, when $p>1$, the two first terms are subdominant, and at leading order in $\gamma$ we have
\begin{eqnarray}
    \tau_\gamma(x_0) \approx \frac{2\, \gamma^{1 -\frac{2}{p}}}{v_0^2} \int_0^{x_0\, \gamma^{\frac{1}{p}}}d\tilde z\, e^{\frac{2\alpha}{v_0^2} \, z^p} \int_{\tilde z}^{+\infty} d\tilde y \, e^{-\frac{2\alpha}{v_0^2}\, \tilde y^p}  \quad \, , \quad \gamma \to + \infty\, .
    \label{taulargegammaintegral}
\end{eqnarray}
To analyse the behavior of this double integral at large $\gamma$, we need to distinguish the cases $p=2$ and $p \neq 2$.   

\begin{figure}[t]
   \centering
    \includegraphics[width=0.5\linewidth]{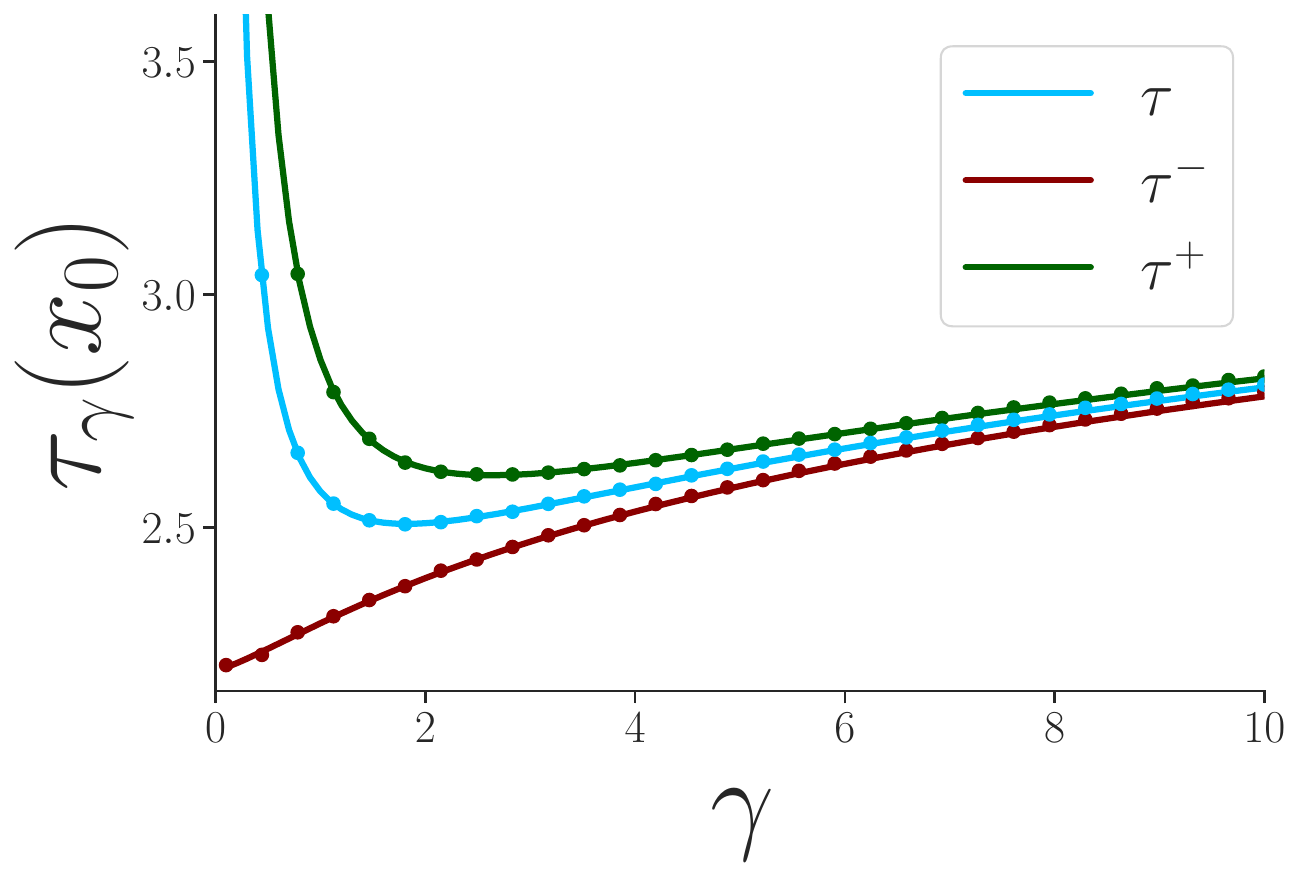}
  \caption{Plot of $\tau_\gamma(x_0), \tau_\gamma^{\pm}(x_0)$ for $V(x)=\mu\, x^2/2$ vs $\gamma$ for fixed $x_0=2.5$ (with $\mu = 1.2$ and $v_0= 1.1$). We clearly see that both $\tau_\gamma(x_0)$ and $\tau^+_\gamma(x_0)$ exhibit a minimum for some optimal tumbling rate. The solid lines correspond to our exact analytical results (\ref{scaling_form})-(\ref{HarmonicscalingF}) while the dots represent results from numerical simulations, showing a perfect agreement.}
  \label{MFPTHarmonicsimugamma} 
\end{figure}

{\bf Case $p=2$}. For the harmonic case where $p=2$ and $\alpha = \mu /2$ it gives
\begin{eqnarray} \label{tau_p2_larg}
    \tau_\gamma(x_0) \approx \frac{2}{v_0^2}\int_0^{x_0\, \sqrt{\gamma}}d\tilde z\, e^{\frac{\mu \tilde z^2}{v_0^2}}\, \int_{\tilde z}^{+\infty}d\tilde y\, e^{-\frac{\mu \tilde y^2}{v_0^2}}  \quad \, , \quad \gamma \to + \infty\, .
\end{eqnarray}
To extract the large $\gamma$ behavior of this integral, we need to characterize the large $\tilde z$ behavior of the integrand. For this purpose we use the asymptotic behavior
\begin{eqnarray}\label{asympt_p2}
\int_{\tilde z}^{\infty}d\tilde y\, e^{-\frac{\mu \tilde y^2}{v_0^2}} = \frac{\sqrt{\pi}\,v_0}{2 \sqrt{\mu}}{\rm erfc}\left(\frac{\sqrt{\mu}\tilde z}{v_0} \right) 
= e^{-\frac{\mu \tilde z^2}{v_0^2}} \left(\frac{v_0^2}{2\mu\tilde z}+ O\left(\frac{1}{\tilde z^2} \right)\right)\, .
\end{eqnarray}
Inserting this asymptotic behavior (\ref{asympt_p2}) in (\ref{tau_p2_larg}) one finds at leading order for large $\gamma$
\begin{eqnarray} \label{tau_p2_largeg}
\tau_\gamma(x_0) \underset{\gamma \to \infty}{\sim} \frac{1}{2\mu} \ln \gamma \;.
\end{eqnarray}
The asymptotic behavior of $\tau_\gamma(x_0)$ for large $\gamma$ (\ref{tau_p2_largeg}) as well as the one for small $\gamma$ (\ref{smallgammabehaviorappendix}) are visible in Fig.~\ref{MFPTHarmonicsimugamma}.

{\bf Case $p\neq2$}. As we study the large $\gamma$ limit of Eq.~(\ref{taulargegammaintegral}), we need to analyse the large $\tilde z$ behavior of the integrand in (\ref{taulargegammaintegral}). For this purpose, let us introduce 
\begin{eqnarray}
    J(\tilde z) = \int_{\tilde z}^{+\infty} d\tilde y \, e^{-\frac{2\alpha}{v_0^2}\tilde y^p} = \int_{\tilde z}^{+\infty} d\tilde y \, e^{-\frac{2\alpha}{v_0^2}\tilde y^p} \left(\frac{2\alpha}{v_0^2}p\, \tilde y^{p-1}\right)\left(\frac{v_0^2}{2\alpha\, p\,  \tilde y^{p-1}} \right) \, ,
\end{eqnarray}
where we have multiplied and divided by $\left(\frac{2\alpha}{v_0^2}p\, \tilde y^{p-1}\right)$ to anticipate an integration by parts. Performing this integration by parts, one gets
\begin{eqnarray} \label{J_IPP}
   J(\tilde z)= \frac{v_0^2}{2 \alpha\, p}\, \tilde z^{1-p}\, e^{-\frac{2\alpha}{v_0^2}\tilde z^p} + (1-p) \frac{v_0^2}{2\alpha\, p}\, \int_{\tilde z}^{+\infty}d\tilde y\, e^{-\frac{2\alpha}{v_0^2}\tilde y^p} \tilde y^{-p}\, .
\end{eqnarray}
One can then applying the same trick to the second integral to show that it behaves as $\propto \tilde z^{1-3p}e^{-\frac{2\alpha}{v_0^2}\tilde z^p}$ such that one can conclude that the leading term in (\ref{J_IPP}) is actually the first one. Hence
from Eq.~(\ref{taulargegammaintegral}) we have
\begin{eqnarray}
    \tau_\gamma(x_0) \underset{\gamma \to \infty}{\sim} \frac{2\, \gamma^{1 -\frac{2}{p}}}{v_0^2} \int_0^{x_0\, \gamma^{\frac{1}{p}}}\, d\tilde z\,  e^{\frac{2\alpha}{v_0^2}\tilde z^p}\,  J(\tilde z)   \quad \, , \quad   J(\tilde z) \underset{\tilde z \to \infty}{\sim} \frac{v_0^2}{2 \alpha\, p}\, \tilde z^{1-p}\, e^{-\frac{2\alpha}{ v_0^2}\tilde z^p} \, .
\label{taufinalAsymptotelargegamma}
\end{eqnarray}
We thus see that we now need to treat separately the case $p>2$ and $p<2$ -- while for $p=2$, we recover the exact same logarithm behavior of Eq.~(\ref{tau_p2_largeg}). 

\begin{enumerate}
    \item $p>2$.
In this case, the integral over $\tilde z$ in  Eq.~(\ref{taufinalAsymptotelargegamma}) converges and therefore the upper-bound of the integral can be sent to $+\infty$. In this case one finds
\begin{eqnarray}
    \tau_\gamma(x_0) \underset{\gamma \to \infty}{\sim} A_p\, \gamma^{1 -\frac{2}{p}}\, ,
\label{taufinalAsymptotelargegamma}
\end{eqnarray}
where $A_p$ is a constant 
\begin{eqnarray}
    A_p = \frac{1}{v_0^2} \int_0^{+\infty}\, d\tilde z\,  e^{\frac{2\alpha}{v_0^2 }\tilde z^p}\,  J(\tilde z) \, .
\label{taufinalAsymptotelargegamma}
\end{eqnarray}
The MFPT is thus diverging as $\gamma \to \infty$. 

\item $1<p<2$

Here, the integral over $\tilde z$ in  Eq.~(\ref{taufinalAsymptotelargegamma}) diverges as $\gamma \to \infty$ and one gets
\begin{eqnarray}
    \tau_\gamma(x_0) \sim \frac{\gamma^{1 -\frac{2}{p}}}{\alpha\, p} \int_0^{x_0\, \gamma^{\frac{1}{p}}}\, d\tilde z\, \tilde z^{1-p} \sim \frac{\gamma^{1 -\frac{2}{p}}}{ \alpha\, p} \frac{1}{2-p}\left(x_0\, \gamma^{\frac{1}{p}}\right)^{2-p}\, .
\label{taufinalAsymptotelargegamma}
\end{eqnarray}
Therefore, the MFPT converges to a constant

\end{enumerate}
\begin{eqnarray}
    \tau_\gamma(x_0) \underset{\gamma \to \infty}{\sim} \frac{x_0^{2-p}}{\alpha p(2-p)}\, .
\label{taufinalAsymptotelargegamma}
\end{eqnarray}
Hence summarizing, we have
\begin{eqnarray}
\tau_\gamma(x_0) \underset{\gamma \to \infty}{\sim}\begin{cases}
\frac{x_0^{2-p}}{\alpha p(2-p)} \quad {\rm ,} \quad 1<p<2 \\
\\
\frac{1}{2\mu} \ln \gamma \quad {\rm ,} \quad p=2\\
\\
A_p\, \gamma^{1 -\frac{2}{p}} \quad  {\rm ,} \quad p>2
\end{cases}
\, .
\label{var_asymp}
\end{eqnarray}
as given in Eq. (28) in the main text. 

\begin{center}
{\bf C.} {\it Small $x_0$ behavior and optimal tumbling rate $\gamma_{\rm opt}$}
\end{center}

The MFPT in Eq.~(\ref{phase3solappendix}) is well defined when $x_0 =0$ and one has
\begin{equation}
     \tau_\gamma(0) =   \frac{1}{2\gamma}+\int_{0}^{x_e} \frac{dy}{v_0-f(y)} \, \text{exp}\left[\int_0^{y}du\, \frac{2\gamma f(u)}{v_0^2-f(u)^2}\right]\, .
\end{equation}
In fact, it is possible to go to the next order in $x_0$ and find the optimal tumbling rate $\gamma_{\rm opt}$ that minimises the MFPT at small $x_0$. In this case, one expects (as found for instance in the case of the harmonic oscillator $p=2$), that $\gamma_{\rm opt}$ is large in this limit. Therefore we analyse the small $x_0$ behavior of $\tau_\gamma(x_0)$ at large $\gamma$, and we check a posteriori that this assumption is indeed consistent.
From Eq.~(\ref{I2changed}) we obtain the behavior of $I_2(x_0,\gamma)$
\begin{eqnarray}
    I_2(x_0,\gamma) \approx \gamma^{1 -\frac{1}{p}} \frac{2x_0}{v_0^2}\,  K   \quad \, , \quad K = \int_{0}^{+\infty} d\tilde y \, e^{-\frac{2\alpha}{v_0^2}\tilde y^p}\, .
\end{eqnarray}
Regarding $I_1(\gamma)$, Eq.~(\ref{I1changedlimit}) gives
\begin{eqnarray}
    I_1(\gamma) \approx \gamma^{-\frac{1}{p}}\, \frac{K}{v_0}\, .
\end{eqnarray}
Hence, we have
\begin{eqnarray}
    \tau_\gamma(x_0) \approx \frac{1}{2\gamma} +\gamma^{-\frac{1}{p}}\, \frac{K}{v_0}+\gamma^{1 -\frac{1}{p}} \frac{2x_0}{v_0^2}\,  K  \underset{p>1}{\approx} \gamma^{-\frac{1}{p}}\, \frac{K}{v_0}+\gamma^{1 -\frac{1}{p}} \frac{2x_0}{v_0^2}\,  K \, .
\end{eqnarray}
Using Eq.~(3) of the main text as well as $x_e=(v_0/(\alpha\, p))^{1/(p-1)}$, we have $\gamma_c = v_0(p-1)/x_e$ and we finally write
\begin{eqnarray}
    \tau_\gamma(x_0) \approx \frac{1}{\gamma_c}\left(A_\gamma+B_\gamma \, \frac{x_0}{x_e}\right)\, ,
\end{eqnarray}
with
\begin{eqnarray}
A_\gamma \approx \kappa_p \left( \frac{\gamma_c}{\gamma} \right)^{\frac{1}{p}} \quad, \quad B_\gamma \approx 2(p-1) \kappa_p \left(\frac{\gamma_c}{\gamma} \right)^{\frac{1}{p}-1} \;,
\end{eqnarray}
and
\begin{eqnarray}
    \kappa_p= \frac{K}{v_0}\, \left(\frac{x_e}{v_0(p-1)}\right)^{1+\frac{1}{p}}\, .
\end{eqnarray}
One can then extract the optimal tumble rate 
\begin{eqnarray}
    \gamma_{\rm opt} = \frac{\gamma_c}{2\, (p-1)^2}\, \frac{x_e}{x_0} \quad \, , \quad x_0 \ll 1.
\end{eqnarray}
In the particular case $p=2$ corresponding to the harmonic potential, we have $\gamma_{\rm opt} \sim \frac{v_0}{2x_0}$.

\begin{center}
{\bf D.} {\it Large $x_0$ and optimal tumbling rate $\gamma_{\rm opt}$}
\end{center}

\begin{figure}[t]
   \centering
    \includegraphics[width=0.5\linewidth]{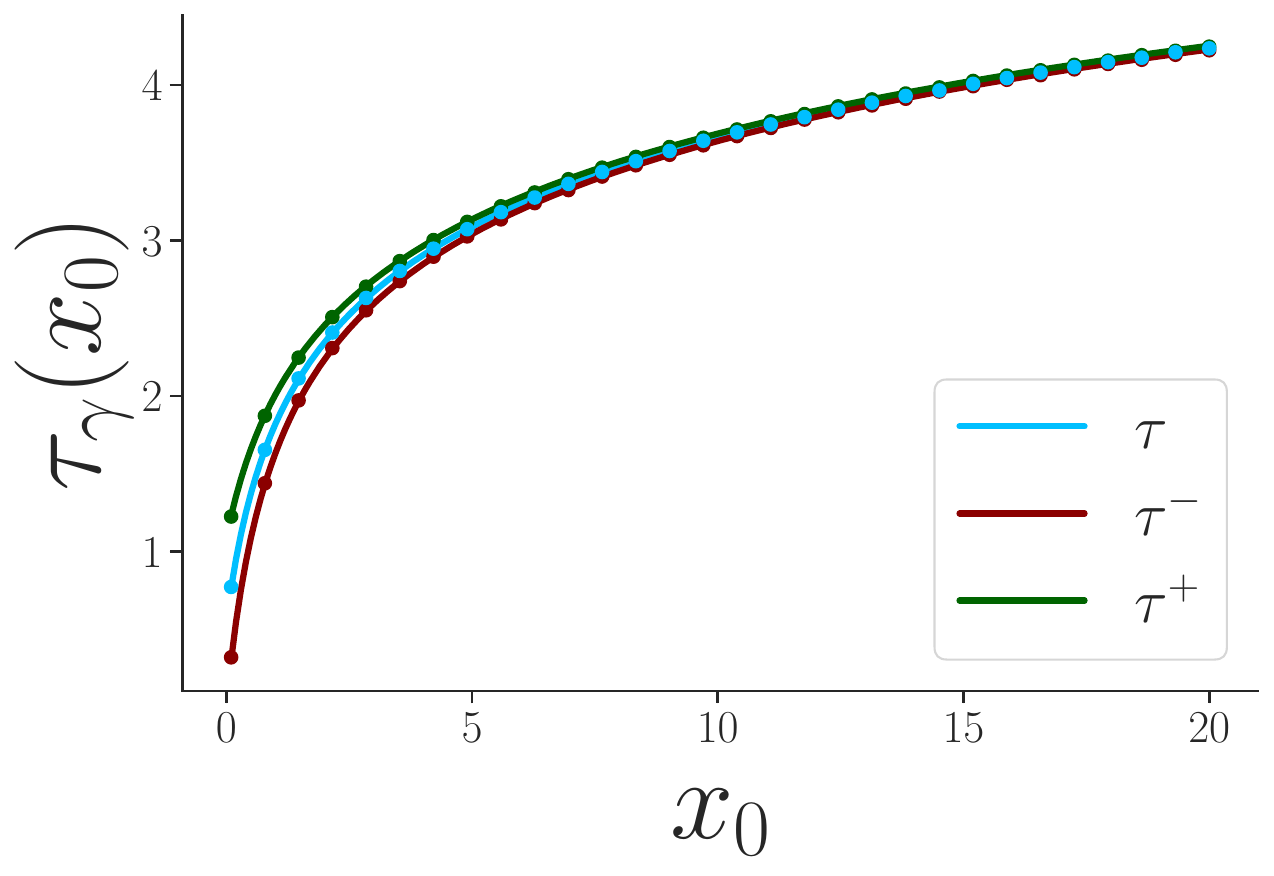}
  \caption{Plot of $\tau_\gamma(x_0), \tau_\gamma^{\pm}(x_0)$ for $V(x)=\mu\, x^2/2$ vs $x_0$ for fixed $\gamma = 2.5$ (with $\mu = 1.2$ and $v_0= 1.1$). The solid lines correspond to our exact analytical results in Eqs. (\ref{tauharmonicscaling})-(\ref{tauharmonicscaling2}) while the dots represent results from numerical simulations, showing a perfect agreement.}
  \label{MFPTHarmonicsimux0} 
\end{figure}

\noindent {\bf Harmonic case $p=2$.} When considering a harmonic potential $V(x)=\mu\, x^2 / 2$ leading to a force $f(x) = -\mu\, x$, it is possible to explicitly compute the integrals in Eq. (18) of the main text and write the MFPT in the scaling form
\begin{eqnarray}
    \tau(\mu, \beta,u) = \frac{1}{\mu} \, \mathcal{F}\left(\beta = \frac{\gamma}{\mu},u = \frac{\mu x_0}{v_0}\right)\, ,
    \label{tauharmonicscaling}
\end{eqnarray}
with
\begin{eqnarray}\label{tauharmonicscaling2}
        &\mathcal{F}\left(\beta,u\right) = \frac{\sqrt{\pi}}{2\beta} \frac{\Gamma\left(1+\beta\right)}{\Gamma\left(\frac{1}{2}+\beta\right)}\left[1 + 2\beta\, u \, {}_2 F_1\left(\frac{1}{2},1 +\beta,\frac{3}{2},u^2\right)\right] \nonumber\\
&-  (2\beta + 1) \, \frac{u^2}{2} \, {}_3 F_2\left(\{1,1,\frac{3}{2}+\beta\};\{\frac{3}{2},2\};u^2\right)\, ,
\end{eqnarray}
where ${}_2 F_1(\cdot;z)$ and ${}_3 F_2(\cdot;z)$ are hypergeometric functions \cite{Grad}. At large $u$, the asymptotic behaviors of the hypergeometric functions can be obtained and one gets at leading order
\begin{equation}
    \mathcal{F}\left(\beta,u\right) = \ln u + \frac{1}{2}\, H_{\beta-\frac{1}{2}} + \frac{\sqrt{\pi}\, \Gamma(\beta)}{2\, \Gamma\left(\frac{1}{2}+\beta\right)} + \ln 2 + o(1)\, ,
\end{equation}
where $H_x=\int_0^1 dx\, \frac{1-t^x}{1-t}$ is the analytic continuation of the $n^{th}$ harmonic number, where $n$ is a positive integer. This asymptotic behavior for large $x_0$ is visible in Fig. \ref{MFPTHarmonicsimux0}. The optimal adimensional rate $\beta^* = \gamma_{\rm opt}/\mu$ can be numerically calculated by cancelling the partial derivative of $\mathcal{F}\left(\beta,u\right)$ with respect to $\beta$. It is solution of
\begin{eqnarray}
    \frac{\sqrt{\pi}}{\Gamma\left(\beta^*+\frac{1}{2}\right)}\left[\psi_0(\beta^*)-\psi_0\left(\beta^*+\frac{1}{2}\right)\right] + \psi_1\left(\beta^*+\frac{1}{2}\right) = 0\, ,
\end{eqnarray}
where $\psi_m(z)$ is the polygamma function of order $m$. Solving numerically this equation gives $\beta^* \approx 1.38657\dots$, as given in the main text in Eq. (26).

\vspace*{0.5cm}
\noindent{\bf The case $p>2$.} Let us first study the function $I_2$
\begin{eqnarray}
    I_2(x_0,\gamma) = \int_0^{x_0}dz\, \frac{1}{v_0^2-f(z)^2} \int_{x_e}^{z} dy \left(f'(y)- 2\gamma \right) \text{exp}\left[\int_y^{z}du\, \frac{-2\gamma f(u)}{v_0^2-f(u)^2}\right]\, .
\end{eqnarray}
When $p>2$, $\frac{-2\gamma f(u)}{v_0^2-f(u)^2} \sim u^{1-p}$ thus the integral inside the exponential converges and one can write 
\begin{eqnarray} \label{largez_p>2}
    I_2(x_0,\gamma) = \int_0^{x_0}dz\, \frac{\text{exp}\left[\int_{+\infty}^{z}du\, \frac{-2\gamma f(u)}{v_0^2-f(u)^2}\right]}{v_0^2-f(z)^2} \int_{x_e}^{z} dy \left(f'(y)- 2\gamma \right) \text{exp}\left[\int_y^{+\infty}du\, \frac{-2\gamma f(u)}{v_0^2-f(u)^2}\right]\, .
\end{eqnarray}
At large $z$ one can easily see that the integral over $y$ is dominated by the term $f'(y)$ and therefore this integral behaves as $f(z) \propto z^{p-1}$. Therefore, the integrand (as a function of $z$) in (\ref{largez_p>2}) behaves as $\propto z^{1-p}$ such that if $p>2$ then $I_2(x_0,\gamma)$ converges to a constant as $x_0 \to \infty$, which reads 
\begin{eqnarray}
    \lim_{x_0 \to \infty} \tau_\gamma(x_0) &{=}& {\tilde{d}_\gamma = } \frac{1}{2\gamma}+\int_{0}^{x_e} \frac{dy}{v_0-f(y)} \, \text{exp}\left[\int_0^{y}du\, \frac{2\gamma f(u)}{v_0^2-f(u)^2}\right]  \\
     &+&\int_0^{+\infty}dz\, \frac{1}{v_0^2-f(z)^2} \int_{x_e}^{z} dy \left(f'(y)- 2\gamma \right) \text{exp}\left[\int_y^{z}du\, \frac{-2\gamma f(u)}{v_0^2-f(u)^2}\right]  \quad \, , \quad p > 2\, ,
    \end{eqnarray}
One can indeed check numerically that, for $f(x) = - \alpha p |x|^{p-1}$, the constant $\tilde{d}_\gamma$, as a function of $\gamma$, admits a minimum $\gamma_{\rm opt}$ that depends on $p$. 

\vspace*{0.5cm}
\noindent{\bf The case $1<p<2$.} One first rewrites, using intergation by parts, 
\begin{eqnarray}
- 2\gamma \int_{x_e}^{z} dy\,  \text{exp}\left[\int_y^{z}du\, \frac{-2\gamma f(u)}{v_0^2-f(u)^2}\right] = \frac{f(z)^2-v_0^2}{f(z)} + \int_{x_e}^z dy\, \left(\frac{v_0^2}{f(y)}-f(y)\right)'\, \text{exp}\left[\int_y^{z}du\, \frac{-2\gamma f(u)}{v_0^2-f(u)^2}\right]\, ,
\end{eqnarray}
such that
\begin{eqnarray}
    I_2(x_0,\gamma) =-\int_0^{x_0}dz\, \frac{1}{f(z)} +\int_0^{x_0}dz\, \frac{1}{v_0^2-f(z)^2} \int_{x_e}^{z} dy \left(\frac{v_0^2}{f(y)}\right)' \text{exp}\left[\int_y^{z}du\, \frac{-2\gamma f(u)}{v_0^2-f(u)^2}\right]\, .
\end{eqnarray}
With $f(x) = -\alpha\, p\, x^{p-1}$, it simplifies to 
\begin{eqnarray}
    I_2(x_0,\gamma) = \frac{x_0^{2-p}}{\alpha\, p\, (2-p)} +\int_0^{x_0}dz\, \frac{1}{v_0^2-f(z)^2} \int_{x_e}^{z} dy \frac{v_0^2(p-1)}{\alpha\, p} \, y^{-p}\text{exp}\left[\int_y^{z}du\, \frac{-2\gamma f(u)}{v_0^2-f(u)^2}\right]\, .
\label{largexsmallpI2analysis}
\end{eqnarray}
We now show that the integral is convergent such that one can take the limit $x_0 \to \infty$ for the upper boundary of the integral. An integration by part gives
\begin{eqnarray}
   \int_{x_e}^{z} dy\, y^{-p}\, \text{exp}\left[\int_y^{z}du\, \frac{-2\gamma f(u)}{v_0^2-f(u)^2}\right] = \frac{v_0^2-f(z)^2}{2\gamma\, z^p f(z)} - \int_{x_e}^{z}\, dy \, \frac{d}{dy}\left(\frac{1}{y^p}\frac{v_0^2-f(y)^2}{2\gamma\, f(y)}\right)\, \text{exp}\left[\int_y^{z}du\, \frac{-2\gamma f(u)}{v_0^2-f(u)^2}\right]\, .
\end{eqnarray}
At large $y$, for $1<p<2$, $\frac{d}{dy}\left(\frac{1}{y^p}\frac{v_0^2-f(y)^2}{2\gamma\, f(y)}\right)\sim -\frac{\alpha^2\, p^2}{2\gamma}\frac{1}{y^2} \ll \frac{1}{y^p}$ such that the second term is sub-leading. The integrand in Eq.~(\ref{largexsmallpI2analysis}) thus behaves as $\sim z^{1-2p}$ such that the integral converges (for $1<p<2$). Therefore one has
\begin{eqnarray}
     \lim_{x_0 \to \infty} \tau_\gamma(x_0) -\frac{x_0^{2-p}}{\alpha\, p\, (2-p)}  &{=}& d_\gamma =   \int_0^{+\infty}dz\, \frac{1}{v_0^2-f(z)^2} \int_{x_e}^{z} dy \frac{v_0^2(p-1)}{\alpha\, p} \, y^{-p}\text{exp}\left[\int_y^{z}du\, \frac{-2\gamma f(u)}{v_0^2-f(u)^2}\right]\\
     &+&  \frac{1}{2\gamma}+\int_{0}^{x_e} \frac{dy}{v_0-f(y)} \, \text{exp}\left[\int_0^{y}du\, \frac{2\gamma f(u)}{v_0^2-f(u)^2}\right] \,    \quad \, , \quad 1<p < 2\, ,
    \end{eqnarray}
One can indeed check numerically that, for $f(x) = - \alpha p |x|^{p-1}$, the constant $d_\gamma$, as a function of $\gamma$, admits a minimum $\gamma_{\rm opt}$ that depends on $p$.

\begin{center}
{\bf IV. Mean first passage time in the presence of a reflecting wall}
\end{center}

In the presence of a unique stable negative turning point, we have shown that the MFPT $\tau_\gamma(x_0)$ is given by Eq.~(18) of the main text.
In this section we consider instead the case where the force is such that $|f(x_0)|<v_0$, and we calculate the MFPT for $x_0\in [0,L]$ when there is a reflecting wall at $L$ as was done in \cite{RTPPSG}\footnote{Note that there the target is placed in $X>0$, while the initial position is to the left of the target $x<X$.}. Let us denote by $T_\gamma(x_0,L)$ the MFPT in the presence of this reflecting wall. For $L \to \infty$, it allows us to retrieve the MFPT for a linear potential (given in Eq.~(32) in the main text). We also compare our formula (18) for $\tau_\gamma(x_0)$ with the one obtained for $T_\gamma(x_0,L)$.

\begin{center}
{\bf A.} {\it Case of $|f(x)|<v_0$ with a reflecting wall at $x=L$}
\end{center}

What happens if the particle starts exactly at the wall $x(0)=L$? The constraint that we need to impose is that if the net velocity of the particle $f(L) \pm v_0$ is positive, it has to stay at the wall, i.e., $x(dt)=L$ (since this is a hard wall). Hence we need to distinguish two cases:
%
\begin{itemize}
    \item \textit{Case 1}: $f(L) + v_0 < 0$: here the velocity of the particle is negtive in both states and it moves towards the origin -- just as if there were no wall at $x=L$, i.e.,
\begin{equation}
x(dt) = \begin{cases}
L + \left[f(L) + v_0\right]dt &\text{, with probability } 1 - \gamma\, dt \text{ and } \sigma(t+dt) = +1 \\
L+ \left[f(L) - v_0\right]dt&, \text{with probability } \gamma\, dt \text{ and } \sigma(t+dt) = -1 \;.
\end{cases}
\end{equation}
In this case, the dynamical evolution of $Q^{\pm}(L,t)$ is not affected by the wall, as given in Eqs. (4)-(5) of the main text.

    \item \textit{Case 2}: $f(L) + v_0 > 0$. The particle has positive speed in the '+' state and tries to cross the reflecting wall, while it is negative in the '-' state and moves normally towards the origin. Thus the dynamics reads 
\begin{equation}
x(dt) = \begin{cases}
L &\text{, with probability } 1 - \gamma\, dt \text{ and } \sigma(t+dt) = +1 \\
L+ \left[f(L) - v_0\right]dt&, \text{with probability } \gamma\, dt \text{ and } \sigma(t+dt) = -1
\end{cases}
\end{equation}
    As a consequence the backward Fokker-Planck equations are modified at the wall. It leads to
\begin{equation}
    Q^+(L,t+dt)= (1-\gamma\, dt)Q^+(L,t) + \gamma \, dt Q^-(L+\left[f(L) - v_0\right]dt,t)\, \, ,
\end{equation}
where we recall that $Q^\pm$ are the survival probabilities in the states $\sigma = \pm1$.
We therefore have the following differential equation
\begin{equation}
    \partial_t Q^+(L,t) = -\gamma\, Q^+(L,t) +\gamma\, Q^-(L,t)\, . \label{FPattheWall}
\end{equation}

If we compare it to the Fokker-Planck equation for the survival probability of an RTP -- Eq. (4) in the main text -- we see that we have to impose
\begin{equation}
    \theta(f(L)+v_0)\, \partial_{x_0}Q^+(x_0,t)\Big|_{x_0=L}=0\, .
\label{reflectivecondition}
\end{equation}
\end{itemize}
We conclude that when $f(L)> -v_0$, Eq. (\ref{reflectivecondition}) applies and we can impose 
\begin{equation}
    \partial_{x_0}Q^+(x_0,t)\Big|_{x_0=L}=0\, .
\end{equation}
In other words, as we have
\begin{equation}
T_+(x_0,L) =  -\int_0^{+\infty} dt\, t\,  \partial_t Q^+(x_0,t)\, ,
\end{equation}
it is equivalent to impose
\begin{equation} \label{BC}
\partial_{x_0}T_\gamma^+(x_0,L)\Big|_{x_0=L}=0\, .
\end{equation}

The positive and negative MFPT obey 
\begin{equation}
\left[v_0^2 -f(x_0)^2\right] \partial^2_{x_0} T_\gamma^\pm + \left[2\gamma f(x_0) - f(x_0)f'(x_0) \pm v_0f'(x_0)\right]\partial_{x_0} T_\gamma^\pm = -2\gamma\, .
\end{equation}
Solving the equation for $T^+_\gamma(x_0,L)$, together with the boundary condition (\ref{BC}), one finds
%
%
We obtain $T_\gamma^+$ after another integration
\begin{equation}
    T_\gamma^+(x_0,L)= \int_{0}^{x_0}dz\, \frac{2\gamma}{v_0+f(z)}\, \int_{z}^{L} \frac{dy}{v_0-f(y)} \frac{H(z)}{H(y)} + B\, .
    \label{reflectivetau+appendix}
\end{equation}
To fix $B$, we impose $T_\gamma^-(0,L) = 0$, and for this purpose we find $T_\gamma^-$ using
\begin{equation}
    \left[v_0 + f(x_0)\right] \partial_{x_0}T_\gamma^+(x_0,L) - \gamma T_\gamma^+(x_0,L) + \gamma T_\gamma^-(x_0,L)= -1 \, ,
\end{equation}
so that
\begin{equation}
    \left[v_0 + f(x_0)\right] \left(\frac{2\gamma}{v_0+f(x_0)}\, \int_{x_0}^{L} \frac{dy}{v_0-f(y)} \frac{H(x_0)}{H(y)}\right) - \gamma \left(\int_{0}^{x_0}dz\, \frac{2\gamma}{v_0+f(z)}\, \int_{z}^{L} \frac{dy}{v_0-f(y)} \frac{H(z)}{H(y)} + B\right) + \gamma T_\gamma^-= -1 \, .
\end{equation}
Taking the limit $x_0 \to 0$, we obtain
\begin{equation}
    2\gamma\, \int_{0}^{L} \frac{dy}{v_0-f(y)} \frac{1}{H(y)} - \gamma B = -1\, ,
\end{equation}
i.e.,
\begin{equation}
B = \frac{1}{\gamma} +2\, \int_{0}^{L} \frac{dy}{v_0-f(y)} \frac{1}{H(y)} \, .
\end{equation}
Finally, we obtain
\begin{equation}
    T_\gamma^+(x_0,L)= \frac{1}{\gamma} +2\, \int_{0}^{L} \frac{dy}{v_0-f(y)} \frac{1}{H(y)}+ \int_{0}^{x_0}dz\, \frac{2\gamma}{v_0+f(z)}\, \int_{z}^{L} \frac{dy}{v_0-f(y)} \frac{H(z)}{H(y)}\, ,
\end{equation}
as well as
\begin{equation}
T_\gamma^-(x_0,L) =2\, \int_{0}^{L} \frac{dy}{v_0-f(y)} \frac{1}{H(y)}  -2\, \int_{x_0}^{L} \frac{dy}{v_0-f(y)} \frac{H(x_0)}{H(y)}+ \int_{0}^{x_0}dz\, \frac{2\gamma}{v_0+f(z)}\, \int_{z}^{L} \frac{dy}{v_0-f(y)} \frac{H(z)}{H(y)}   \, .
\end{equation}
The 'average' MFPT $T_\gamma(x_0) = 1/2(T_\gamma^+ + T_\gamma^-)$ is given by
\begin{equation} \label{final_T_L}
\begin{split}
       T_\gamma(x_0,L) &= \frac{1}{2\gamma}+2\, \int_{0}^{L} \frac{dy}{v_0-f(y)} \frac{1}{H(y)} -\, \int_{x_0}^{L} \frac{dy}{v_0-f(y)} \frac{H(x_0)}{H(y)}\\
        &+ \int_{0}^{x_0}dz\, \frac{2\gamma}{v_0+f(z)}\, \int_{z}^{L} \frac{dy}{v_0-f(y)} \frac{H(z)}{H(y)}   \, .
\end{split}
\end{equation}

\vspace*{0.3cm}

\begin{center}
{\bf B.} {\it Proof of the identity $\tau_\gamma(x_0) = T_\gamma(x_0,L=x_e)$}
\end{center}

Quite remarkably, for $V(x) = \alpha |x|^p$ with $p>1$, we show here that our formula (\ref{tau_FP_1}), for $x_0\leq x_e$, is related to $T_\gamma(x_0,L) $ via $\tau_\gamma(x_0) = T_\gamma(x_0,L=x_e)$. In addition, we show that $T_\gamma(x_0,L=x_e)$ is also well defined for $x_0 > L=x_e$ and still coincides with $\tau_\gamma(x_0)$ in (\ref{tau_FP_1}). This identity, for $x_0 < x_e$, can be qualitatively understood by noting that the edge $x_e$, seen from the left $0\leq x_0 < x_e$, behaves effectively like a reflecting wall (see Fig. 3 in the main text). In addition, on that interval $0\leq  x_0 < x_e$, one has $|f(x)|< v_0$ -- which is then a situation similar to the one studied in Ref. \cite{RTPPSG} with indeed $L = x_e$.

Setting $L=x_e$ in Eq. (\ref{final_T_L}), one finds
\begin{equation}
\begin{split}
       T_\gamma(x_0, L=x_e) &= \frac{1}{2\gamma}+2\, \int_{0}^{x_e} \frac{dy}{v_0-f(y)} \frac{1}{H(y)} -\, \int_{x_0}^{x_e} \frac{dy}{v_0-f(y)} \frac{H(x_0)}{H(y)}\\
        &+ \int_{0}^{x_0}dz\, \frac{2\gamma}{v_0+f(z)}\, \int_{z}^{x_e} \frac{dy}{v_0-f(y)} \frac{H(z)}{H(y)}   \, ,
\end{split}
\label{taureflective}
\end{equation}
where we recall that $H(x_0) = \text{exp}\left[\int_0^{x_0}dx \frac{-2\gamma f(x)}{v_0^2-f(x)^2}\right]$. On the other hand, solving the full ODE (\ref{generalequationontau}) for $\tau_\gamma$ with the condition (\ref{conditionTauplus}) at the stable negative fixed point $x_e$ we have shown 
\begin{equation}
\begin{split}
     \tau_\gamma(x_0) =  - \frac{1}{2\gamma v_0}\left[f(0) + \int_{0}^{x_e} dy \frac{f'(y)- 2\gamma}{H(y)}\right]  +\int_0^{x_0}dz\, \frac{H(z)}{v_0^2-f(z)^2} \int_{x_e}^{z} dy \frac{f'(y)- 2\gamma}{H(y)}\, .
\end{split}
\label{tau1TPbis}
\end{equation}
Here we prove that these two formulae (\ref{taureflective}) and (\ref{tau1TPbis}) are the same.

To prove that they do coincide, we will first prove that their derivative with respect to $x_0$ are the same. First the derivative of Eq. (\ref{taureflective}) is
\begin{equation}
\begin{split} \label{T}
    \partial_{x_0} T_\gamma(x_0,x_e) &= \frac{1}{v_0 -f(x_0)} - \int_{x_e}^{x_0} \frac{dy}{v_0-f(y)}\, \frac{2\gamma\, f(x_0)}{v_0^2-f(x_0)^2}\, \frac{H(x_0)}{H(y)} + \frac{2\gamma}{v_0 + f(x_0)} \int_{x_0}^{x_e} \frac{dy}{v_0-f(y)} \, \frac{H(x_0)}{H(y)}\\
    &= \frac{1}{v_0 -f(x_0)} + 2\gamma\, \frac{v_0}{v_0^2 - f(x_0)^2} \int_{x_0}^{x_e} \frac{dy}{v_0-f(y)} \, \frac{H(x_0)}{H(y)}\, .
\end{split}
\end{equation}
On the other hand, if we derive Eq. (\ref{tau1TPbis}) with respect to $x_0$ we obtain
\begin{equation}
\begin{split}
    \partial_{x_0} \tau_\gamma(x_0)& = \frac{1}{v_0^2-f(x_0)^2} \int_{x_e}^{x_0} dy \left(f'(y)- 2\gamma \right)\,  \frac{H(x_0)}{H(y)}\\
    &    =\frac{-2\gamma}{v_0^2-f(x_0)^2} \int_{x_e}^{x_0} dy \,  \frac{H(x_0)}{H(y)} + \frac{1}{v_0^2-f(x_0)^2} \int_{x_e}^{x_0} dy \, f'(y)\,  \frac{H(x_0)}{H(y)}\, .
\end{split}
\label{tauistaustep1}
\end{equation}
Now we perform an integration by part in the second term of the last equality in Eq. (\ref{tauistaustep1}). We note $v=f(y)$, $v'=f'(y)$ and $u=H(x_0)/H(y)$ such that $u'=2\gamma f(y)/(v_0^2 -f(y)^2)\, H(x_0)/H(y)$. Also remembering that we have $f(x_e)=-v_0$, we find
\begin{equation} 
\begin{split}
    \partial_{x_0} \tau_\gamma(x_0)& =\frac{-2\gamma}{v_0^2-f(x_0)^2} \int_{x_e}^{x_0} dy \,  \frac{H(x_0)}{H(y)} + \frac{1}{v_0^2-f(x_0)^2}\left[f(y)\,  \frac{H(x_0)}{H(y)}\right]_{x_e}^{x_0} - \frac{1}{v_0^2-f(x_0)^2} \int_{x_e}^{x_0} dy \, \frac{2\gamma f(y)^2}{v_0^2-f(y)^2} \, \frac{H(x_0)}{H(y)}\\
    & = -2\gamma \frac{v_0^2}{v_0^2-f(x_0)^2}\int_{x_e}^{x_0} dy\, \frac{1}{v_0^2 - f(y)^2}\, \frac{H(x_0)}{H(y)} + \frac{f(x_0)}{v_0^2-f(x_0)^2}+\frac{v_0}{v_0^2 -f(x_0)^2}\, \frac{H(x_0)}{H(x_e)}\, .
\end{split}
\label{tauistaustep2}
\end{equation}
Subtracting the two expressions (\ref{tauistaustep2}) and (\ref{T}), we obtain
\begin{equation}
    \begin{split}
        \partial_{x_0} T_\gamma(x_0) - \partial_{x_0} \tau_\gamma(x_0) &=\frac{1}{v_0 -f(x_0)} - \frac{f(x_0)}{v_0^2-f(x_0)^2}-\frac{v_0}{v_0^2 -f(x_0)^2}\, \frac{H(x_0)}{H(x_e)}\\
        &+2\gamma \frac{v_0}{v_0^2 - f(x_0)^2} \int_{x_0}^{x_e}dy\, \left[\frac{1}{v_0 - f(y)}- \frac{v_0}{v_0^2 - f(y)^2}\right]\, \frac{H(x_0)}{H(y)}\\
         &=\frac{v_0}{v_0^2-f(x_0)^2}-\frac{v_0}{v_0^2 -f(x_0)^2}\, \frac{H(x_0)}{H(x_e)} +2\gamma \frac{v_0}{v_0^2 - f(x_0)^2} \int_{x_0}^{x_e}dy\, \frac{f(y)}{v_0^2 - f(y)^2}\, \frac{H(x_0)}{H(y)}\\
         &=\frac{v_0}{v_0^2-f(x_0)^2}-\frac{v_0}{v_0^2 -f(x_0)^2}\, \frac{H(x_0)}{H(x_e)} +\frac{v_0}{v_0^2 - f(x_0)^2}\, \left[\frac{H(x_0)}{H(y)}\right]_{x_0}^{x_e}=0\, .
    \end{split}
\end{equation}
To conclude, we now show that $T_\gamma(x_0=0,x_e) = \tau_\gamma(0)$. First, one has
\begin{equation} \label{T_01}
\begin{split}
        T_\gamma(0,x_e) &= \frac{1}{2\gamma}+\int_{0}^{x_e} \frac{dy}{v_0-f(y)} \frac{1}{H(y)}\, ,
\end{split}
\end{equation}
while $\tau_\gamma(x_0=0)$ is given by
\begin{equation}
\begin{split}
    \tau_\gamma(0) &=  - \frac{1}{2\gamma v_0}\left\{f(0) + \int_{0}^{x_e} dy \left(f'(y)- 2\gamma\right)\frac{1}{H(y)}\right\}\\
     &= - \frac{1}{2\gamma v_0}\left\{f(0) -2\gamma\, \int_0^{x_e} \frac{dy}{H(y)} + \left[f(y)\frac{1}{H(y)}\right]_0^{x_e} -2\gamma \int_0^{x_e}dy\, \frac{f(y)^2}{v_0^2-f(y)^2}\, \frac{1}{H(y)}\right\}\\
     &= - \frac{1}{2\gamma v_0}\left\{\frac{-v_0}{H(x_e)} -2\gamma \int_0^{x_e}dy\, \left[1+\frac{f(y)^2}{v_0^2-f(y)^2}\right]\, \frac{1}{H(y)}\right\}\\
     &= - \frac{1}{2\gamma v_0}\left\{\frac{-v_0}{H(x_e)} -2\gamma \int_0^{x_e}dy\, \frac{v_0^2}{v_0^2-f(y)^2}\, \frac{1}{H(y)}\right\}\, . \label{T_02}
\end{split}
\end{equation}
Now we subtract the two expressions (\ref{T_01}) and (\ref{T_02}) and we get
\begin{equation}
\begin{split}
       T_\gamma(0) - \tau_\gamma(0) &= \frac{1- \frac{1}{H(x_e)}}{2\gamma} + \int_0^{x_e}dy\, \left[\frac{1}{v_0-f(y)}-\frac{v_0}{v_0^2-f(y)^2}\right]\frac{1}{H(y)} \\
       &= \frac{1- \frac{1}{H(x_e)}}{2\gamma} + \int_0^{x_e}dy\, \left[\frac{f(y)}{v_0^2-f(y)^2}\right]\frac{1}{H(y)}\\
       &= \frac{1- \frac{1}{H(x_e)}}{2\gamma} + \left[\frac{1}{2\gamma\, H(y)}\right]_0^{x_e}=0\, ,
\end{split}
\end{equation}
Hence $T_\gamma(x_0,L=x_e)=\tau_\gamma(x_0)$, as announced in the main text. 

\end{document}